\documentclass[aps,pre,twocolumn,groupedaddress]{revtex4-2}

\usepackage{graphicx}
\usepackage{dcolumn}
\usepackage{bm}
\usepackage{caption}
\usepackage{subcaption}
\usepackage{siunitx}
\usepackage{natbib}

\begin{document}


\title{Role of the ratio of tangential to normal stiffness coefficient on the behaviour of vibrofluidised particles}
\author{Alok Tiwari}
\email{alok.tiwari@iitb.ac.in}
\author{Sourav Ganguli}%
\affiliation{%
Department of Energy Science and Engineering, Indian Institute of Technology Bombay, Mumbai 400076, India\\
}%
\author{V. Kumaran}%
\affiliation{%
	Department of Chemical Engineering, Indian Institute of Science Bangalore, Bengaluru 560012, India
	}%
\author{Manaswita Bose}%
 \email{manaswita.bose@iitb.ac.in}
\affiliation{%
Department of Energy Science and Engineering, Indian Institute of Technology Bombay, Mumbai 400076, India\\
}%


\begin{abstract}

The selection of parameters in the contact law for inter-particle interactions affects the results of simulations of flowing granular materials. The present study aims to understand the effect of the ratio of tangential to normal spring stiffness coefficient ($\kappa$) on inter-particle contact behaviour in terms of the rotational coefficient of restitution determined using data obtained from multi-particle simulations. The effect of $\kappa$ on the profiles of the micro- and macroscopic properties of particles in a vibrofluidised bed is also investigated. The Discrete Element Method (DEM) is used to simulate a vertically vibrated fluidised bed using the open-source software LAMMPS. The inter-particle and wall-particle contact forces are determined using the linear spring-dashpot (LSD) model. The distribution of the mean co-ordination number, force during the contact, contact regimes, and rotational coefficient of restitution are determined from the data obtained from simulations. It was shown that $\kappa$ plays a significant role in the distribution of inter-particle contacts between different regimes and, thereby, the velocity distribution and profiles of statistically averaged properties of the vibrofluidised particles. Our results show that for particles with surface friction coefficient $\mu>0.1$, the commonly used value $\kappa=\frac{2}{7}$ results in quantitatively different results from those obtained using $0.67 \le \kappa < 1$, a range consistent with the realistic values of Poisson ratios for simple materials.


\end{abstract}

\maketitle

\section{\label{sec:Intro}Introduction}
Realistic modelling of the inter-particle interaction and selection of parameters are crucial for the accurate simulation of the macroscopic behaviour of real granular materials \cite{Johnson1987, Silbert2001, Silbert2002, Guo2015}.
Traditionally, inter-particle or wall-particle interactions are modelled either considering the collision to be ``instantaneous'', known as the hard sphere model (HS) \cite{Rapaport1980} or representing the deformation of the contact surface with an appropriate combination of a spring (linear or non-linear), a dashpot and a slider. Cundall and Strack \cite{Cundall1979} were the first to use a combination of a linear spring, dashpot, and a slider to model the contact between colliding particles. Their results show that the behaviour depends on the ratio of the normal to the tangential spring stiffness constant and the Poisson ratio. Later, Tsuji and coworkers \cite{Tsuji1992} implemented Hertz and Mindlin's no-slip models for determining the normal and the tangential contact forces, respectively\cite{Hertz1882}. Silbert \textit{et al.} simulated the flow of granular materials over an inclined plane using linear spring dashpot and Hertz-Mindlin model \cite{Silbert2001}. They have performed simulations for the normal spring stiffness constant $k_n \approx 10^5 mg/d$, where $m$ and $d$ are the mass and diameter of individual particles. They selected the ratio of the normal ($k_n$) to the tangential ($k_t$) spring stiffness constant $(\kappa=(k_t/k_n))$ as $(2/7)$ since the time-periods of contacts for this value are equal \cite{Schafer1996,Thornton2011}; however, a selection of $\kappa = (2/7)$ also leads to mutually exclusive sticking and sliding regimes \cite{Walton1993}. Silbert \textit{et al.} \citep{Silbert2001} have suggested that the selection of $\kappa$ has a marginal effect on the average behaviour of the granular flow over the inclined plane. They have also shown that the velocity and the volume fraction profiles on an inclined plane converge for normal spring constant $k_n > 10^5 mg/d$. In a series of later articles, the effect of spring stiffness constant on Bagnold scaling \cite{Brewster2008}, coordination number \cite{Reddy2010}, and normal stress \cite{Ketterhagen2005} is investigated. In most of the earlier and many recent articles \cite{Bharathraj2017, Kim2020, Debnath2022}, $\kappa$ is maintained as $(2/7)$. However, Thornton and coworkers \cite{Thornton2011} explicitly recommended the selection of the ratio in the range $0.67 \le \kappa < 1$, based on the realistic values of Poisson ratio, $0 < \nu \leq 0.5$. Their work is limited to single-particle contact. Since $\kappa=(2/7)$ is mostly used in earlier investigations, it is important to study other values of $\kappa$ to examine whether there is any qualitative difference.

Loading and unloading behaviour of a single particle in contact with a planar surface or with another particle has been investigated in the literature using experiments \citep{Cole2007}, a discrete element approach,  modelling the contact using linear and Hertz spring \cite{Thornton2009}, finite element method \cite{Rathbone2015} and smooth particle hydrodynamics \cite{Vyas2023}. Tangential and rotational coefficients of restitution have been determined experimentally \cite{Kharaz2001, Louge2002, TOMAR2018} and from numerical simulations \cite{Thornton2011, Thornton2013}. Recently, Kosinski \textit{et al.} \cite{Kosinski2020} proposed a data-driven approach for developing an empirical relationship expressing the rotational coefficient of restitution as a function of the modulus of elasticity ($Y$), Poisson ratio ($\nu$), normal coefficient of restitution ($e_n$) and impact angle ($\gamma$), $\beta = \beta\left(Y, \nu, e_n, \cot{\gamma}\right)$. However, to our knowledge, complete statistics of the rotational coefficient of restitution from a multi-particle simulation have not yet been reported. 

The present work is motivated by the question of whether the value of $\kappa$ influences the flow behaviour of granular materials.  A vibrofluidised bed is simulated to determine the rotational coefficient of restitution from the position-velocity data of particles obtained from the discrete element method simulations. Vibrofluidisation for this study is chosen because (a) in a vertically vibrated bed, there is no mean linear and angular velocity; (b) unlike a homogeneously cooling system, in a vibrofluidised bed, a steady state can be achieved; (c) volume-fraction (density) and the pressure profiles are linked with the momentum conservation \citep{Kumaran1998}.  
The specific objectives are:
\begin{enumerate}
\item to determine the influence of $\kappa$ on the rotational coefficient of restitution;
\item to investigate the effect of $\kappa$ on the linear and the rotational velocity distribution of particles,  profiles of solid volume fraction, translational and rotational granular temperature, and pressure for different inter-particle coefficients of friction.
\end{enumerate}
To that end, a collection of spherical particles subject to vertical vibration is simulated using the discrete element method. The following sections briefly discuss the simulation methodology, theoretical background, and analysis method.

\section{\label{sec:method}Methodology}

\subsection{Simulation domain}
\begin{figure}[]
\centering
\includegraphics[width=6cm,height=18cm,keepaspectratio]{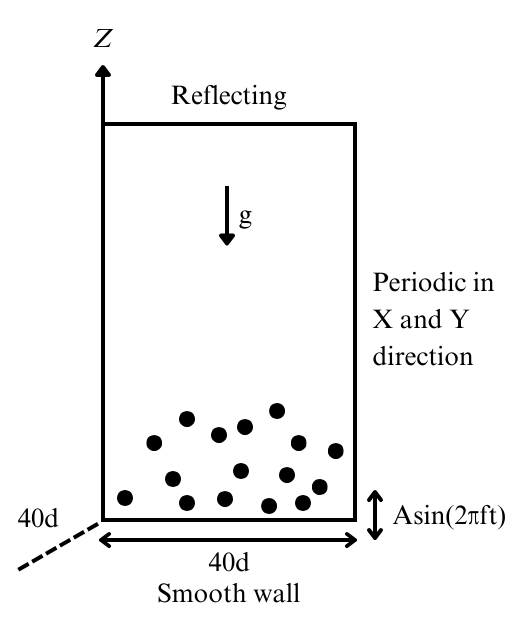}
\caption{Schematic of the simulation domain.} \label{fig:sim_domain}
\end{figure}

Figure ~\ref{fig:sim_domain} shows a schematic of the vertically vibrated fluidised bed. The base plate moves sinusoidally with a velocity $u(t) = 2\pi A f \cos(2\pi f t)$ with a fixed frequency of $100$ Hz and an amplitude 0.3$d$. The top plate is placed at $1000d$ above the base plate and is considered smooth and elastic. The bed is $40d \times 40d$ in cross-section and is periodic in the directions normal to that of the vibration.  A total of 6400 spherical particles are simulated. 
Inter-particle and wall-particle contacts are modelled using a linear spring and a dashpot, popularly known as LSD (Fig. \ref{fig:lsd})\cite{Debnath2022}. Simulations are performed using the open-source software LAMMPS \cite{LAMMPS}.     

\subsection{Discrete element method}\label{sec:DEM}

\begin{figure*}[]
    \subfloat[b][]{\includegraphics[scale=0.25]{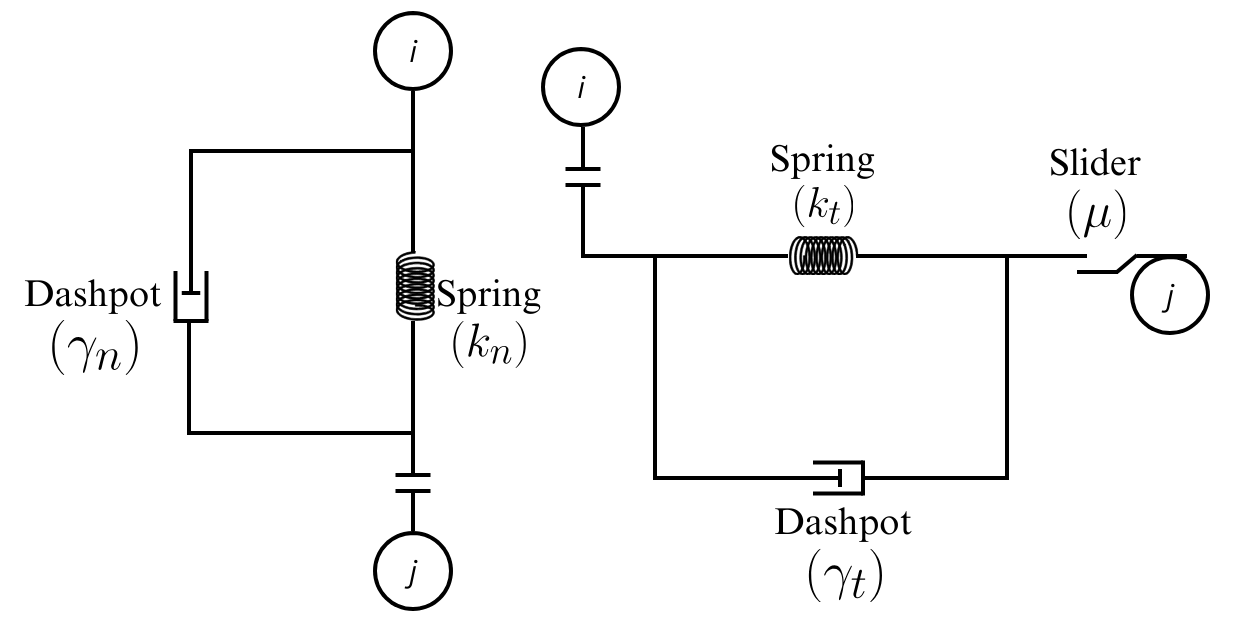}\label{fig:lsd}}
    \subfloat[a][]{\includegraphics[scale=0.25]{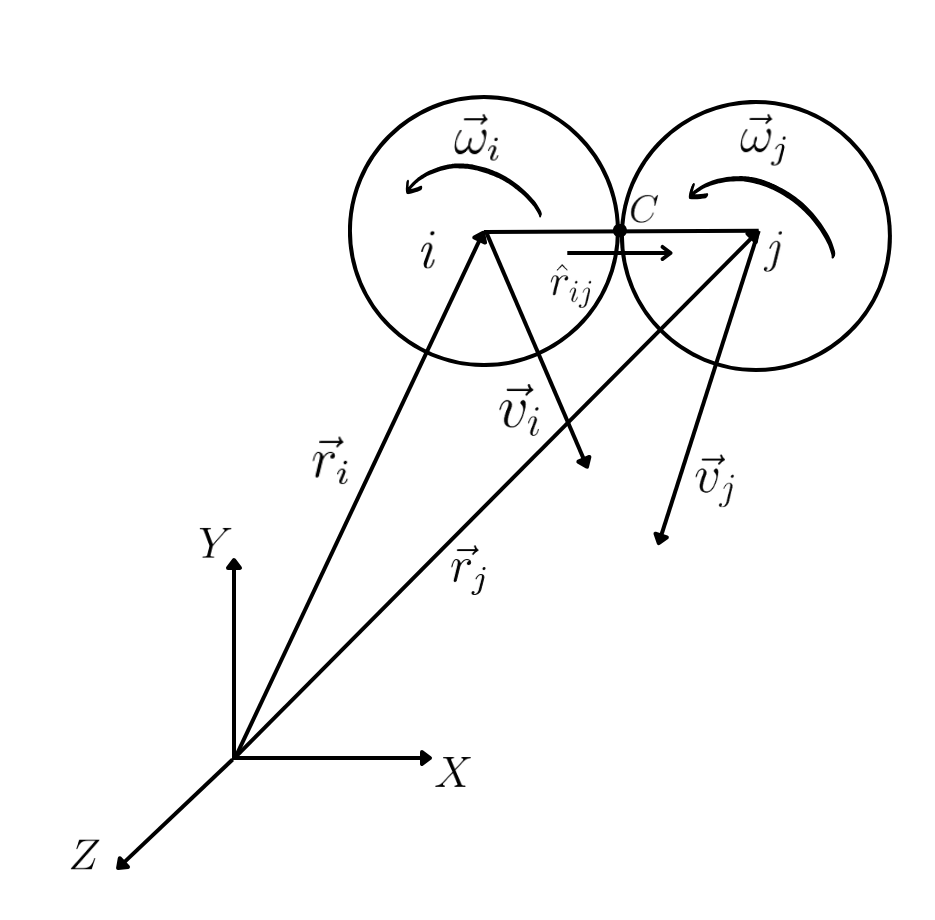}\label{fig:pos_vec}}\\
    \caption{ Schematic representation of (a) the linear spring dashpot model and the (b) hard sphere model
    }
    \label{fig:schematic}
\end{figure*}

The normal ($\vec{F}_{nij}$) and the tangential ($\vec{F}_{tij}$) force exerted on the particle $i$ by particle $j$ at contact, are expressed as:

\begin{equation}\label{eq:normal_force}
\vec{F}_{nij} = -k_n \xi_{n} \hat{r}_{ij} - \gamma_n {\vec{v}_{n}},
\end{equation}

\begin{equation}\label{eq:tangential_force}
\vec{F}_{tij} = - \min \left (k_t \left \| \vec{\xi}_{t}\right \|,\mu  \left \| \vec{F}_{nij} \right \|  \right )\hat{t}_{ij}. 
\end{equation}  
Here, $\xi_{n} = \left \| \vec{r}_{ij} \right \| - d$ is the overlap between $i$ and $j$ particle in the normal direction,
$\hat{r}_{ij}$ is the unit vector joining the centre of particles from $i$ to $j$ (Fig. \ref{fig:pos_vec}), $\gamma_n$ is the damping coefficient in the normal direction, and $\vec{v}_{n} = \left ( \vec{v}_{ij} \cdot \hat{r}_{ij}\right ) \hat{r}_{ij} $ is the velocity of particle $j$ with respect to $i$ in the normal direction. In the present study, the damping in the tangential direction is neglected.  For $\left \| \vec{F}_{tij} \right \| >  \mu  \left \| \vec{F}_{nij} \right \| $, the tangential displacement of the spring is truncated \cite{Kruggel-Emden2008,Luding2008} as $\vec{\xi'}_{t} (t+ \delta t)= - \left (\mu  \left \| \vec{F}_{nij}\right \|/k_t \right) \hat{t}_{ij}$, where $\hat{t}_{ij} = \mbox{}
- \left (\vec{\xi}_{t}/\left \| \vec{{\xi}_{t}} \right \|\right)$ is the unit vector in the direction of tangential displacement at a time $t$.

The normal spring constant
$k_n$ is chosen as $10^{8} (mg/d)$ to ensure that the collisions are mostly binary \cite{Mitarai2003,Reddy2007}. For glass particles of 5 mm diameter, the linear spring stiffness constant is $\approx$ 10$^9 (mg/d)$ \citep{Reddy2007};
however, in the present simulations, $k_n = 10^8 (mg/d)$ is selected to save the computational time without compromising accuracy. Simulations are performed for four different values of $\kappa$, $(2/7, 3/4, 0.85, 0.95)$. 
The normal dashpot coefficient is adjusted using the expression $\gamma_n = \left (2\sqrt{k_nm}\ln e_n \right)/(\sqrt{\pi^2+(\ln e_n)^2})$ to set the inter-particle coefficient of restitution $e_n = 0.95$ \cite{Schafer1996}. The inter-particle friction coefficient $(\mu)$ is varied in the range $0.01$ to $0.5$. The time of contact is estimated as $t_c = \pi \left ( 2k_{n}/m) - (\gamma_n^2/m^2) \right )^{-1/2}$.

Simulations are performed in two steps: (a) First, simulations are run with a time-step $\delta t = t_{c}/10$ for $10^{6}$ time-steps, 
(b) Once the total kinetic energy of the particles is found to reach a steady state, simulations are further run with a lower time-step of $\delta t = t_{c}/100$ to resolve the inter-particle contact. Data is collected to determine the pre- and post-collisional normal and tangential velocities of colliding particles.

\subsection{Background theory for analysis}\label{sec:theory}

Walton \cite{Walton1993} proposed a simple model (Figure ~\ref{fig:pos_vec}) to describe the collision between two spherical particles or a spherical particle and a wall. The two parameters in the model are the normal and the rotational coefficients of restitution, defined as $e_n = -v_n^*/v_n$ and $\beta = - v_s^*/v_s$ respectively \cite{Lun1987, Walton1993}. Here, $v_n$ is the normal component of the relative velocity, and $v_s = \vec{v}_{s} \cdot \hat{k}_{s}$ is the tangential component of the relative surface velocity. The superscript ``asterisk'' denotes the post-collisional quantities. The relative surface velocity ($\vec{v}_{s}$) and the unit vector ($\hat{k}_{s}$) are defined as:

\begin{equation}\label{eq:slip_velocity}
\vec{v}_{s} = \vec{v}_{ij} -  \vec{v}_{n} + \frac{d}{2}\left ( \hat{r}_{ij} \times \vec{\omega}_{ij} \right )
\end{equation}
where $\vec{\omega}_{ij} = \vec{\omega}_i + \vec{\omega}_j$ and,
\begin{equation}
\label{eq:slip_velocity_2}
\hat{k}_{s} = \frac{\vec{v}_{s}}{\left \| \vec{v}_{s} \right \|}.
\end{equation}

Here, $\beta$ varies between -1 and +1, and in general. The contact is defined as a sticking for $0\le \beta \le 1$, and sliding for $-1\le \beta < 0$. In the sliding regime, $ \beta = -1 + (7/2) \mu (1+e_n) (v_{n}/v_{s})$ \cite{Walton1993}.

Walton's assumption that the particle contact can be classified into two mutually exclusive categories of either sticking or sliding may not be physically valid, as experiments and simulations showed that the contact could be partly sticking before the particles slide  \cite{Foerster1994,Kharaz2001, Louge2002, Buck2017, TOMAR2018}. A more realistic classification based on the extended Hertz model was proposed by Maw \textit{et al.} \cite{Maw1980}. They generated plots of  $\psi_2 = (2 \left(1-\nu\right))/(\mu\left(2-\nu\right) (v_s^*/v_n)$ vs. $\psi_1 = (2 \left(1-\nu\right))/(\mu\left(2-\nu\right)) (v_s/v_n)$, where $\nu$ is the Poisson ratio. Here, $\psi_1$ and $\psi_2$ can be interpreted as non-dimensionalized impact and rebound angles, respectively. According to their criterion, for $\psi_1 < 1$, at the beginning of the contact, the surfaces stick to each other and slide at the end of the contact. Contact begins with sliding before two surfaces stick for a short while, and finally slip at the end of the contact for $1 < \psi_1 < 4\,\chi -1$, where $\chi = (7\left(1-\nu\right))/(2\left(2-\nu\right))$ for a sphere. The surfaces slip during the entire contact for $\psi > 4\,\chi -1$.  For collisions that involve gross sliding, where $f_t = - \mu |f_n| \mathrm{sgn} (\vec{v}_s)$ for the entire contact,
$\psi_{2} = \psi_{1} - (7/2) \mu (1+e_n) (v_{n}/v_{s})$. Di Maio and Di Renzo \citep{DiMaio2004} subsequently showed that the conditions of classification of tangential contacts, which were derived based on the Hertz solution, were applicable to results obtained using the linear spring dashpot (LSD) models. They expressed $\chi = 2r\kappa$ where $r= (1+(md^2/4I))^{-1} = 7/2$ for a sphere. 


In the present work, the position and velocity of the particles obtained from the DEM simulations are first analysed to identify the pair of particles in contact, the coordination number (CN) \cite{Reddy2010,Tiwari2022}, and the frequency distribution of CN. The algorithm employed to identify the pair of particles in contact is discussed in Appendix \ref{appendix:algo}. Once the colliding pair of particles is identified, the pre- and post-collision linear and angular velocities are used to obtain $\beta$, $\psi_1$, and $\psi_2$. Using the plots of $\psi_1$ vs. $\psi_2$ and contact force vs. time, different regimes of contact are identified and the criteria \citep{Maw1980} are verified.

\subsection{Analysis of DEM data}\label{sec:Algo}

The instantaneous position $(\vec{r})$, linear $(\vec{v})$ and angular velocities $(\vec{\omega})$ of particles are 
averaged over bins of height equal to one particle diameter to determine the profiles of the ensemble averaged properties such as volume fraction 
\begin{equation}
\varphi = \frac {\pi d^3 \left \langle n_b \right \rangle}{6V_b},
\end{equation}
the granular temperature, both translational $(T)$ and rotational $(T_R)$,
\begin{equation}
T = \frac{1}{2} \left \langle \left \| \vec{V}'\right \|^2 \right \rangle,
\end{equation}
\begin{equation}
 T_R = \frac{2I}{md^2} \left \langle \left \| \vec{\Omega}'\right \|^2 \right \rangle, 
 \end{equation}
 and pressure,
 \begin{equation}
 P =  \frac{6 \varphi u_b{^2} \left \langle \vec{V}' \cdot \vec{V}'  \right \rangle}{\pi d^3 Ng}\ + \frac{d[\vec{F} \cdot \hat{r}_{ij}]} {Ng V_{b}}.
 \end{equation}
 Here, $\vec{V}' = \vec{V} - \left \langle \vec{V} \right \rangle$ and $\vec{\Omega}' = \vec{\Omega} - \left \langle \vec{\Omega} \right \rangle$ are the fluctuating linear and angular velocity of a particle, respectively, $N$ is the total number of particles per unit area, $\langle n_b \rangle $ is the average number of particle centers in a bin, $V_b$ is the bin volume, $g$ is the acceleration due to gravity, and $I= md^2/10$ is the moment of inertia of the spherical particle. $\langle \rangle$ represents the ensemble-averaged properties over more than $6000$ configurations.
 The linear and angular velocities are normalized as $\vec{V} = \vec{v}/{u_b}$ and $\vec{\Omega} = d\vec{\omega}/2u_b$, where $u_{b} = 2 \pi A f$ is the maximum velocity of the vibrating base.   The vertical position of the particle is measured with respect to the vibrating base.   The average translational ($\mbox{KE}_T$)
 and rotational $\mbox{KE}_R$ kinetic energies are defined as,
 \begin{equation}
\mathrm{KE}_T =\frac{1}{2\,h}\int_{0}^{h}\left \langle \left \| \vec{V}'\right \|^2 \right \rangle dz,
 \end{equation}
\begin{equation}
 \mathrm{KE}_R =\frac{2I}{md^2\,h}\int_{0}^{h}\left \langle \left \| \vec{\Omega}'\right \|^2 \right \rangle dz.
 \end{equation}
The time evolution of the kinetic energies is examined to determine whether the system has reached steady state.

\section{Results}\label{sec2}

\subsection{Binary Collisions}

\begin{figure}[]
\centering
\includegraphics[scale=0.25]{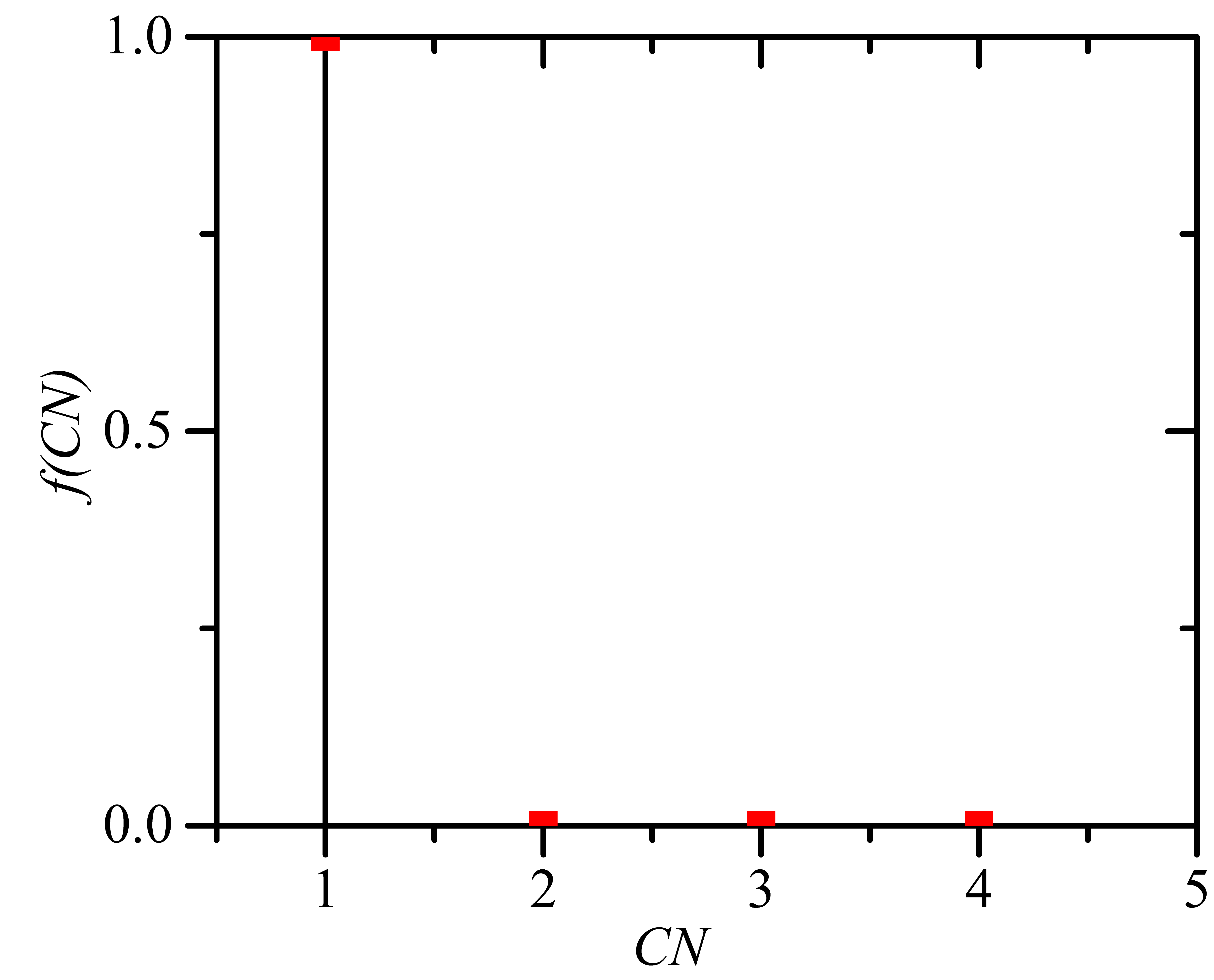}
\caption{Distribution of coordination number (CN) for the colliding particles obtained by tracking individual contacts for $\kappa=3/4$ and $\mu = 0.5$. 
\label{fig:binary_collision}}
\end{figure}

The frequency distribution of the coordination number for $\kappa=3/4$ and $\mu = 0.5$ is plotted in Fig. \ref{fig:binary_collision}. The maximum corresponding to CN=1 implies that the collisions for the selected set of parameters are mostly binary. A similar frequency distribution of CN is observed for other sets of simulations. Once the assumption of binary collision of the hard sphere model is satisfied, further analyses are carried out.

\subsection{Regimes of contact}

\subsubsection{Rotational coefficient of restitution} \label{sec:rot_coeff}

The rotational coefficient of restitution $(\beta = - (v^*_s/v_s))$ is obtained by determining the ratio of post- to pre-collision tangential component of relative surface velocity. Here,
$v_s$ is determined following Equation (\ref{eq:slip_velocity}) by resolving the contacts by $\delta t = t_{c}/100$ (Results for time-step convergence are presented in Appendix \ref{appendix:time_convergence}).
Figure ~\ref{fig:beta_scaling} shows $\beta$ vs. $(7/2)\mu (1+e_n) (v_n/v_s)$ for different values of $\kappa$ and $\mu=0.5$. This plot shows two mutually exclusive regimes for $\kappa=2/7$. In the gross-sliding regime, $\beta$ varies linearly with $(7/2) \mu (1+e_n) (v_n/v_s)$ before it reaches the energy-limiting regime where $\beta = +1$ (within numerical accuracy).  The slip reversal of the contact point in the gross-sliding regime is observed as a consequence of the conservation of the angular momentum. The derivation with a sample example is presented in Appendix \ref{derivation}. A very different behaviour of $\beta$ vs $(7/2) \mu (1+e_n) (v_n/v_s)$ is observed for $\kappa = 3/4, 0.85$ and 0.95. A collapse of data is observed for $(7/2) \mu (1+e_n) (v_n/v_s) \leq 1.3$; beyond this point, $\beta$ passes through a maxima and starts reducing as  $(7/2)\mu (1+e_n) (v_n/v_s)$ increases. The contact becomes near normal as the value of the abscissa increases. It is observed from Figure ~\ref{fig:beta_scaling} that after a certain value of $(7/2) \mu (1+e_n) (v_n/v_s)$, $\beta$ becomes negative, indicating that a sliding effect for the near-normal collision.

Figure \ref{fig:psi2_psi_3_4} shows $\psi_2$ vs $\psi_1$ for $\kappa = 3/4$.  Vertical dotted lines are drawn in Figures ~\ref{fig:beta_scaling} and ~\ref{fig:psi2_psi_3_4} following the criterion proposed by Maw \textit{et al.} \cite{Maw1976} for $\kappa=3/4$. According to this criterion, the cross-over between near-normal and intermediate regimes occurs for $\psi_1 = 1$, which corresponds to $(7/2) \mu (1+e_n) (v_n/v_s) = 5.1$ in Figure ~\ref{fig:beta_scaling}. However, the simulation results show the transition from intermediate to near-normal ($\beta$ becomes negative) at $(7/2) \mu (1+e_n) (v_n/v_s) \approx 6.85$ (which corresponds to $\psi_1 \approx 0.75$). Another method for determining the value of $\psi_1$ for the transition between the near normal and the intermediate regime is by analysing the normal and the tangential forces during contact.

\begin{figure*}[]
    \subfloat[a][]{\includegraphics[scale=0.25]{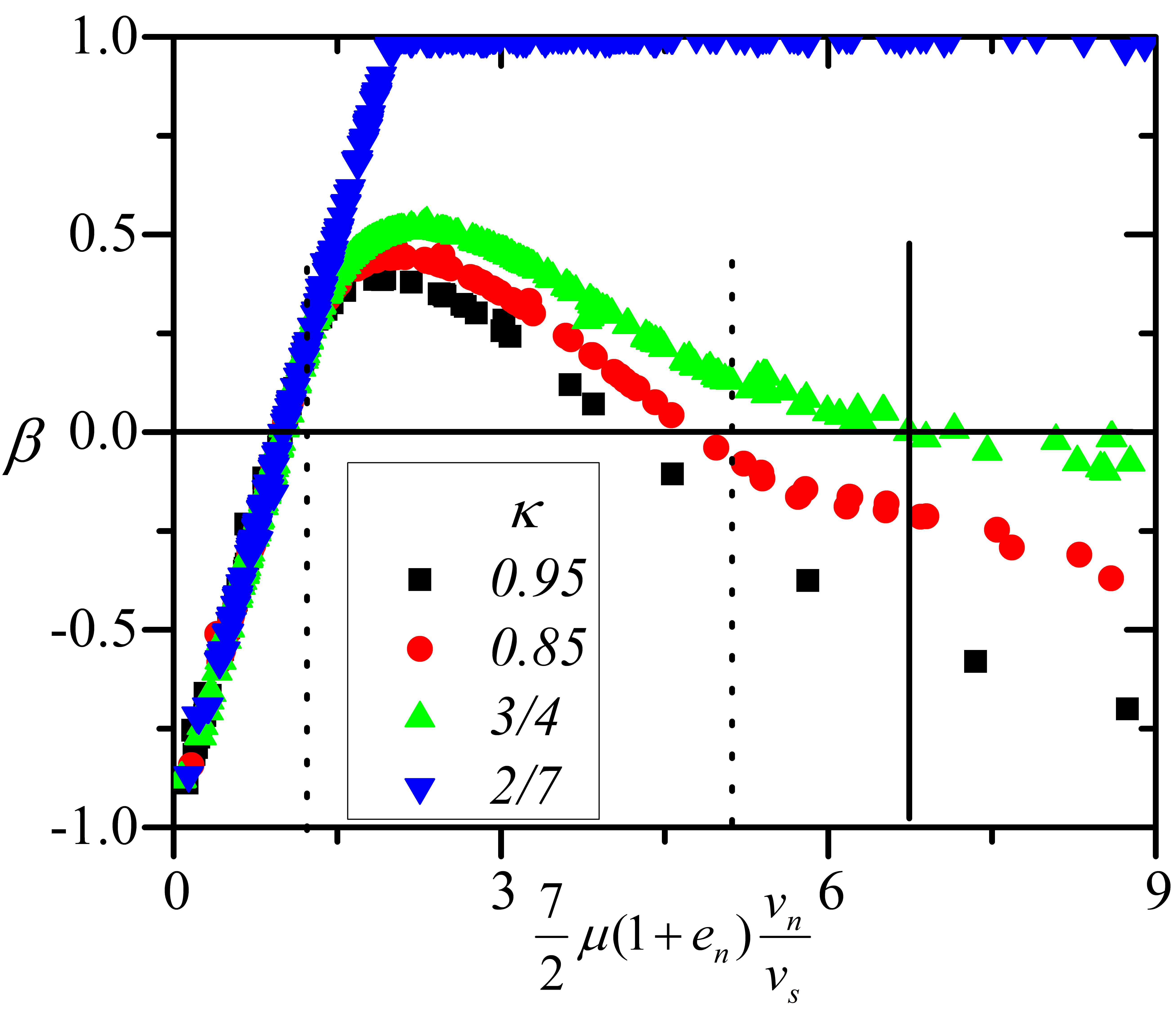}\label{fig:beta_scaling}}
    \subfloat[b][]{\includegraphics[scale=0.25]{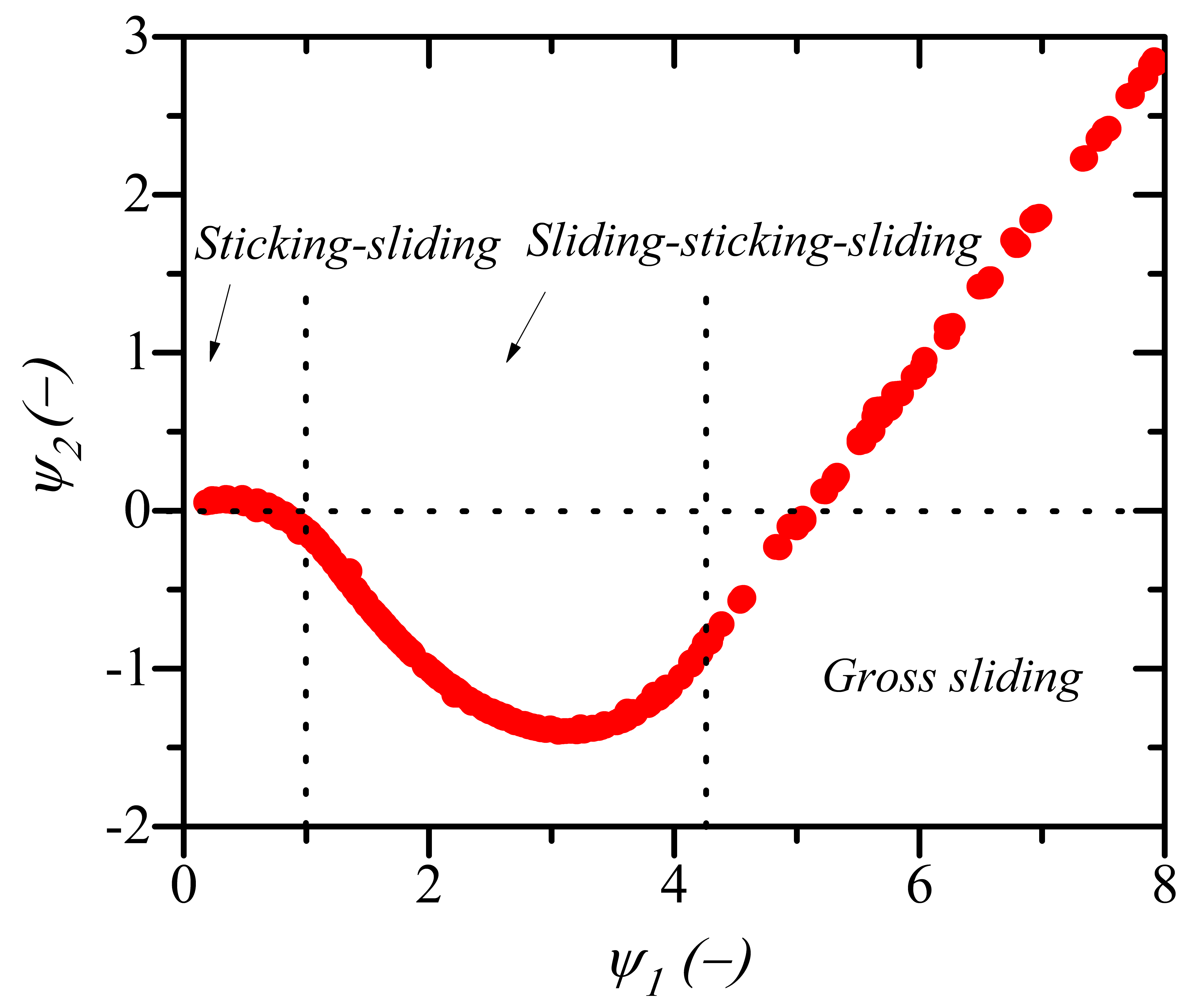}\label{fig:psi2_psi_3_4}}
    \\
    \caption{(a) Plot of rotational coefficient of restitution $(\beta)$ against $(7/2) \mu (1+e_{n}) (v_{n}/v_{s})$ for $\mu=0.5$ and different values of $\kappa$.
    (b) Plot of the non-dimensionalized rebound $(\psi_2)$ and impact angle ($\psi_1$) for $\kappa = 3/4$ and $\mu=0.5$. 
    The vertical dotted line corresponds to the classification criteria given by Maw \textit{et al.} \cite{Maw1976} for transition in tangential contact regime for $\kappa=3/4$. 
    }
    \label{fig:sai_diff_variable}
\end{figure*}


\subsubsection{Normal and tangential force}
  
\begin{figure*}[]
    \subfloat[a][]{\includegraphics[scale=0.25]{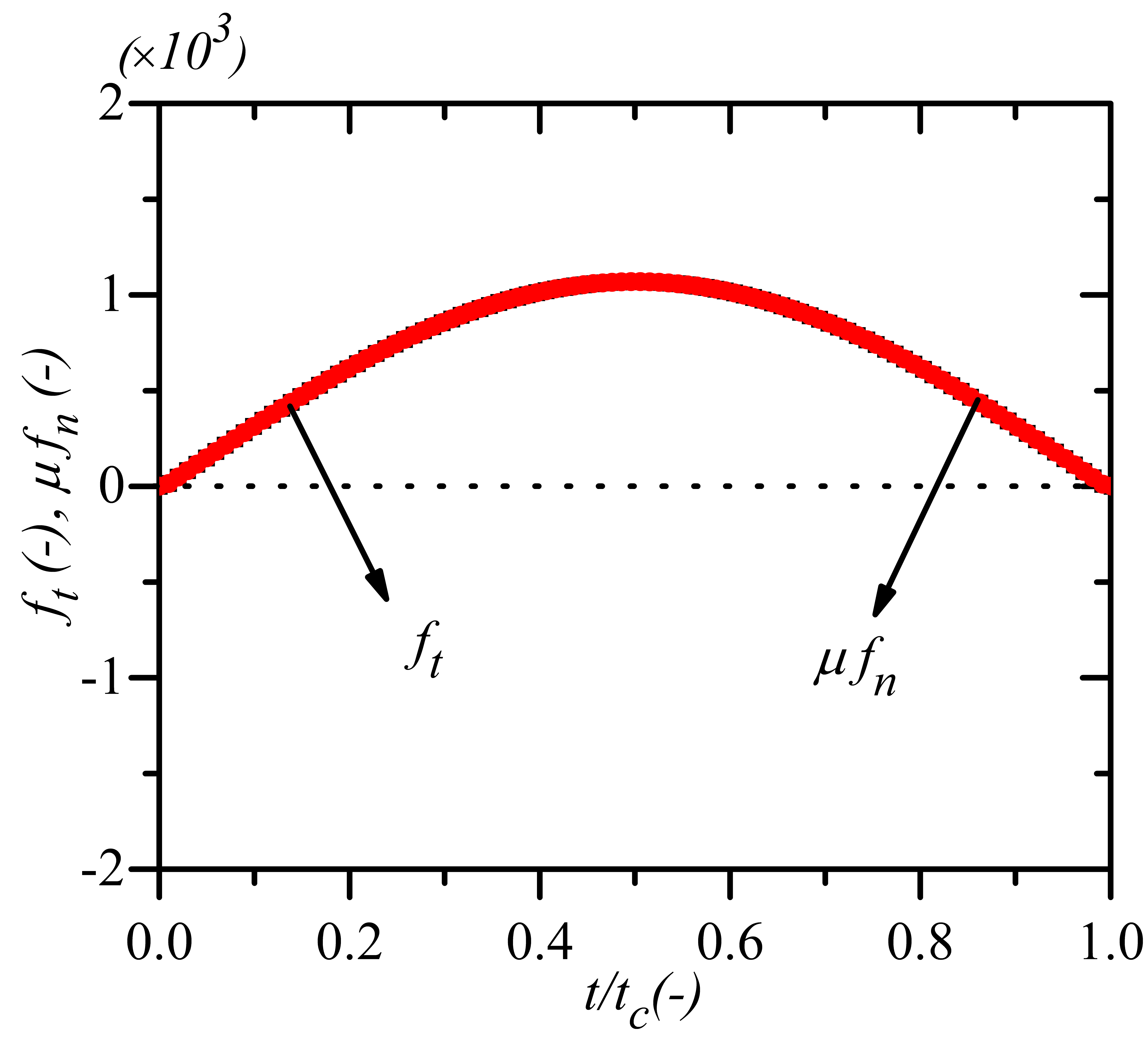}\label{fig:grossslide_3_4}}
    \subfloat[b][]{\includegraphics[scale=0.25]{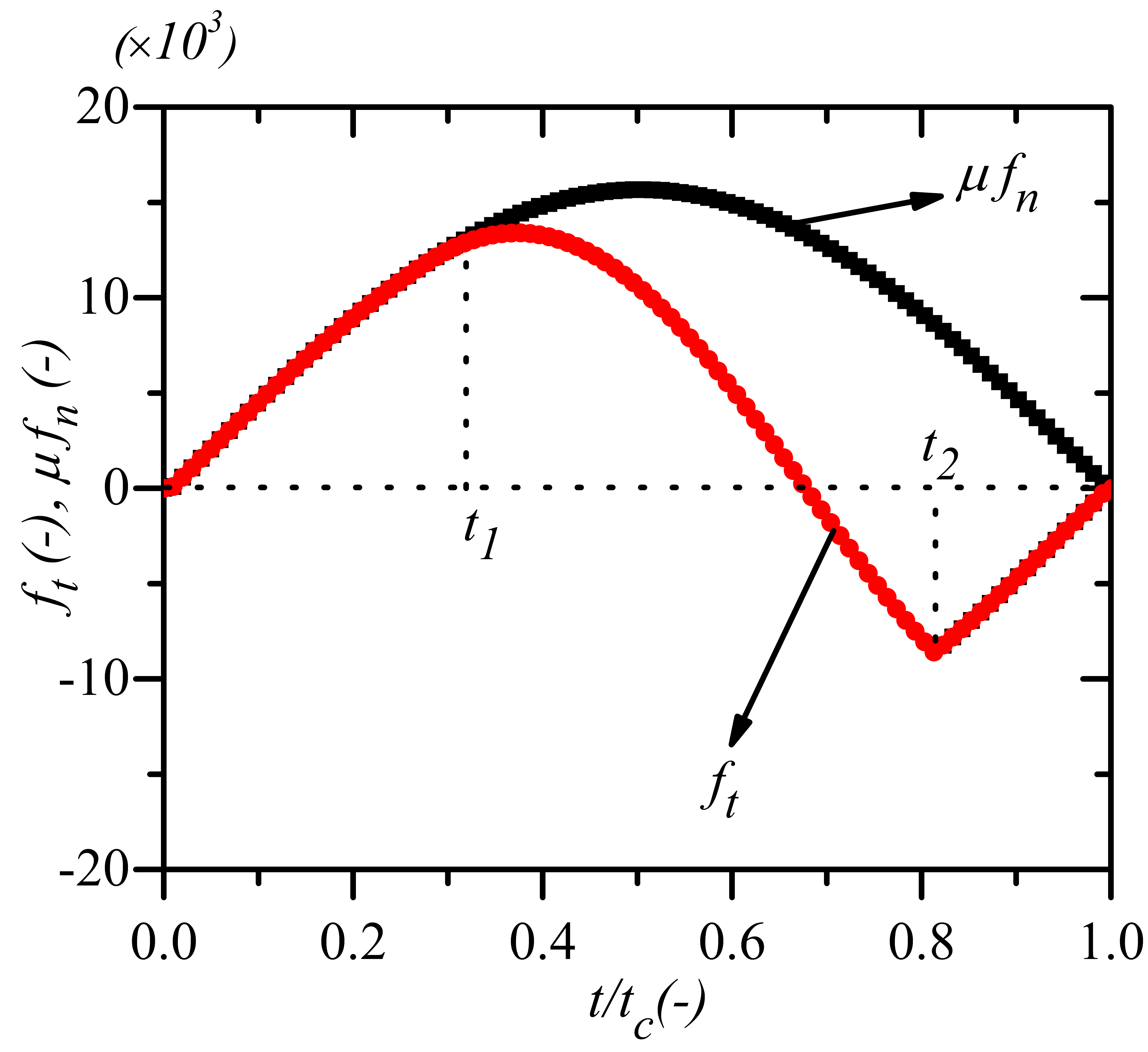}\label{fig:intermed_angle_3_4}}
     \subfloat[c][]{\includegraphics[scale=0.25]{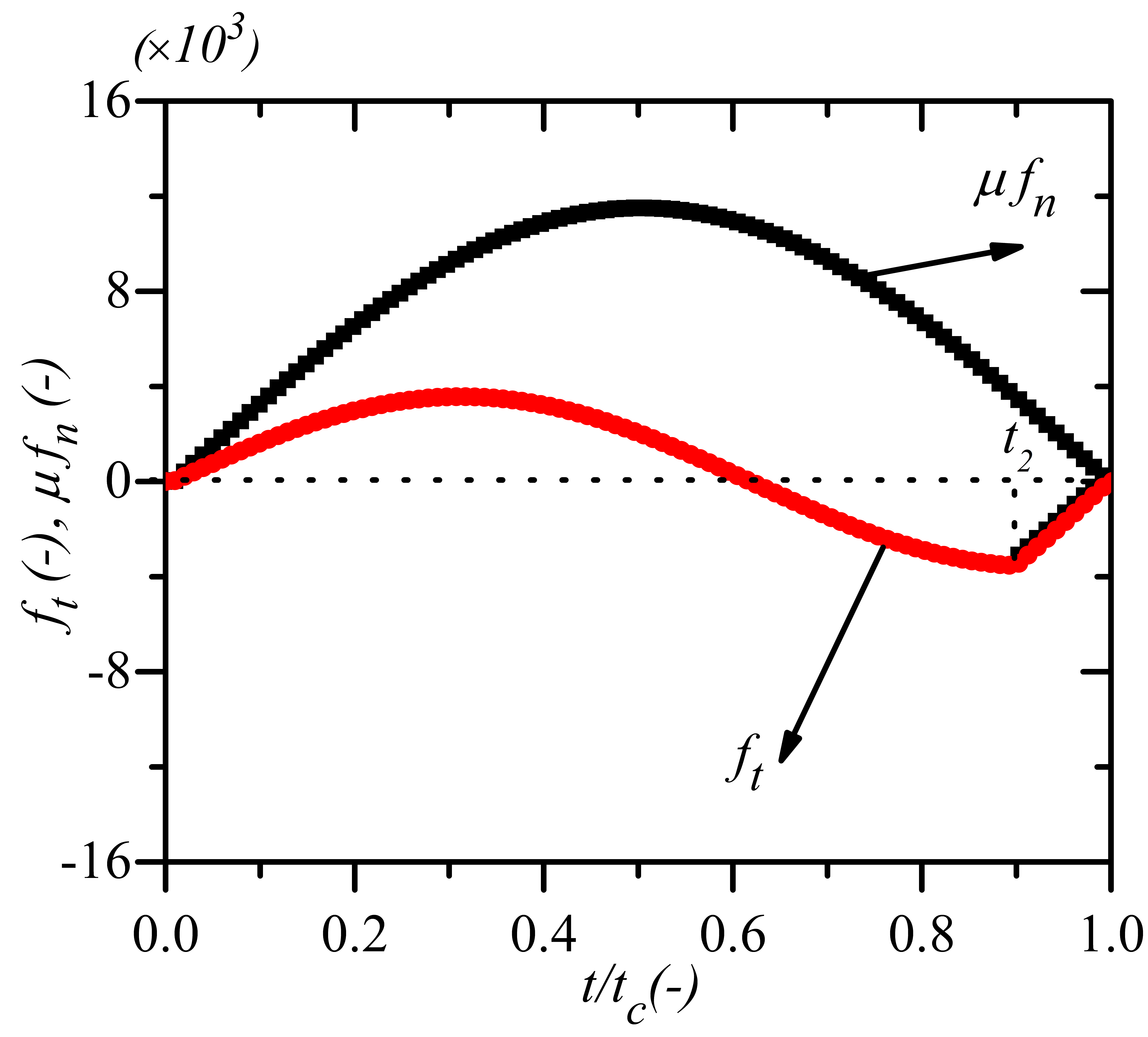}\label{fig:smallimpactsng_3_4}}
         
    \caption{Evolution of $f_t$ and $\mu f_n$ for $\kappa = 3/4$. Plots correspond to (a) gross-sliding at large impact angle $(\psi_1 = 5.0)$ (b) sliding-sticking-sliding at intermediate angle $(\psi_1 = 1.5)$ and (c) sticking-sliding at small impact angle $(\psi_1 = 0.5)$. Here, $f_n = \vec{F}_{ij} \cdot \hat{r}_{ij}$ and $f_{t} = - (\vec{F}_{ij} - f_n \hat{r}_{ij}) \cdot \hat{k}_s$. $f_t$ is multiplied by $-1$ to maintain consistency with Figures of \citep{DiRenzo2004}. The algorithm proposed in \citep{Luding2008} is implemented in LAMMPS for the determination of the tangential component of the force in the intermediate (sliding-sticking-sliding) and near normal (sticking-sliding) regimes.
    \label{fig:3_4_regime}. 
    }
\end{figure*}

Figure \ref{fig:3_4_regime} plots $f_t$ and $\mu f_n$ acting on a particle during a binary contact for three different values of $\psi_1$ representing the gross-sliding, intermediate and near-normal regime and $\kappa=(3/4)$. For $\psi_1 = 5.0$, as the contact is in the gross sliding regime, $f_t = \mu f_n$ during the entire contact (Fig. \ref{fig:grossslide_3_4}). For contact at $\psi_1 = 1.5$, the intermediate regime according to the criterion in \citep{Maw1976}, it is observed that $f_t = \mu f_n$ from the start of the simulation until it reaches time $t_1$ and $f_t < \mu f_n$ from $t_1$ to $t_2$ before sliding till the end of the collision (Fig. \ref{fig:intermed_angle_3_4}). For lower values of $\psi_1$ (say $\psi_1 = 0.5$), i.e., for near normal contact, the tangential force is less than $\mu f_n$ for nearly the entire duration of the contact (Fig. \ref{fig:smallimpactsng_3_4}). Particles slide just before detaching from the contact surface. In this regime, the slip reversal is sometimes not observed, and $\beta$ becomes negative. Similar tangential force-time behaviour is reported in \cite{DiRenzo2004}. 

\begin{figure}[]
\centering
\includegraphics[scale=0.25]{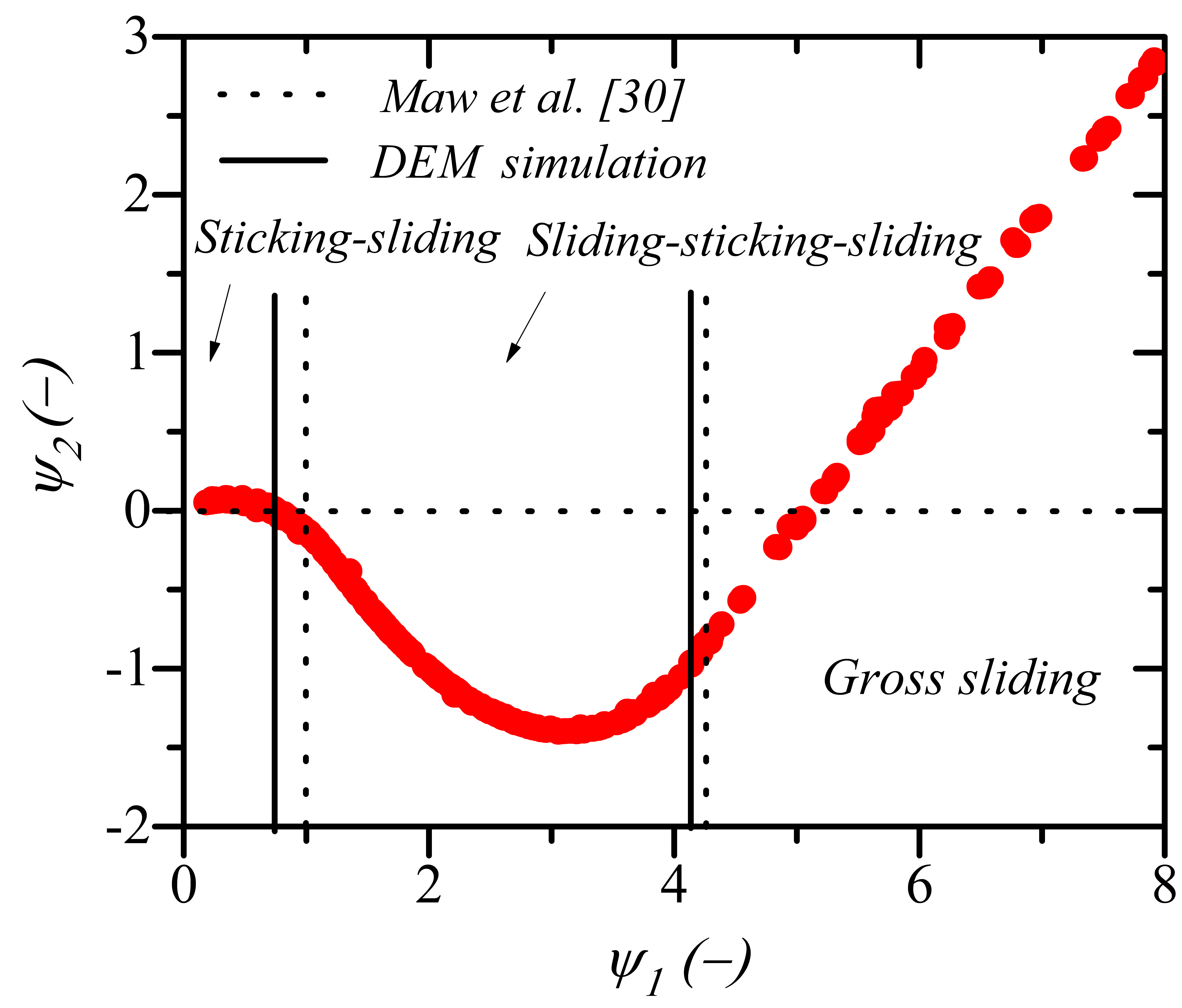}
\caption{ Plot of non-dimensionalized impact $(\psi_2)$ with rebound $(\psi_1)$ angle for $\kappa = 3/4$ and $\mu=0.5$.
\label{fig:psi2_psi1_comparison}
}
\end{figure}

The tangential load versus time plot is used to determine the transition value of $\psi_1$, identifying the near normal and intermediate contact regime. If the initial time during which $f_t = \mu f_n$ is more than $1 \%$ of the total contact time (at the minimum resolution), the collision is considered to be in the intermediate regime. Here, $\psi_1 = 0.5$ (near-normal regime) and $1.5$ (intermediate regime) are chosen as reference values. The final value is obtained using the mid-point search method.  For $\kappa=3/4$, the transition value of $\psi_1$ is found to be $0.718$ (Figure \ref{fig:psi2_psi1_comparison}). Using two different methods, a transition between the near-normal (sticking-sliding) to intermediate (sliding-sticking-sliding) regime is observed at $\psi_1 < 1$.  As $\kappa$ increases, the value of transition $\psi_1$ becomes closer to 1, consistent with the classification of \cite{Maw1976}. Similarly, $\psi_1$ for the transition between intermediate and gross-sliding regimes is determined. It is observed that the value obtained from the simulation agrees well with the proposed criterion of Maw \textit{et al.} \cite{Maw1976} (Figure \ref{fig:psi2_psi1_comparison}). 

\subsubsection{Distribution of contacts}
\begin{figure*}[]
    \subfloat[a][]{\includegraphics[scale=0.25]{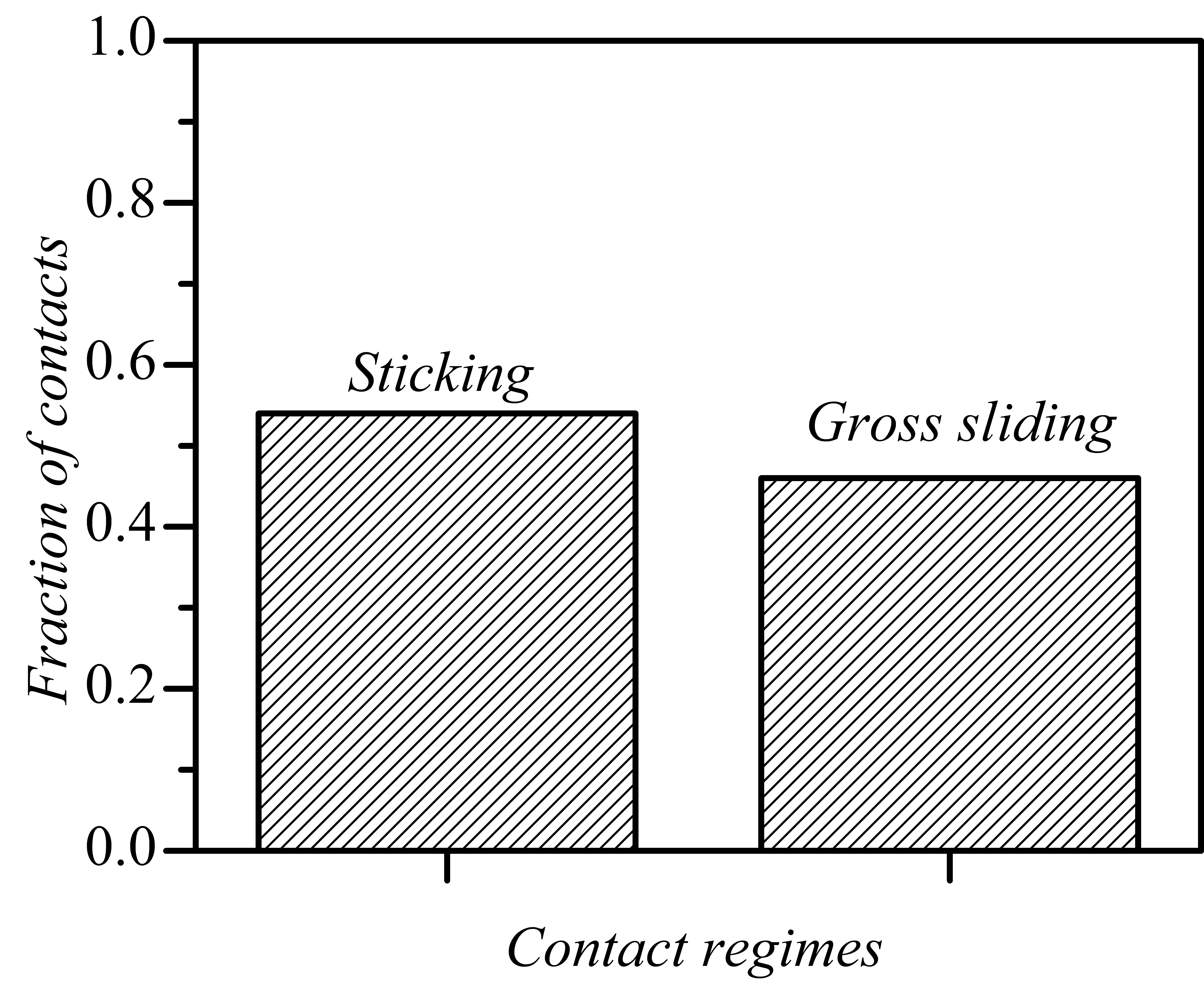}\label{fig:dist_kappa_2_7}}
    \subfloat[b][]{\includegraphics[scale=0.25]{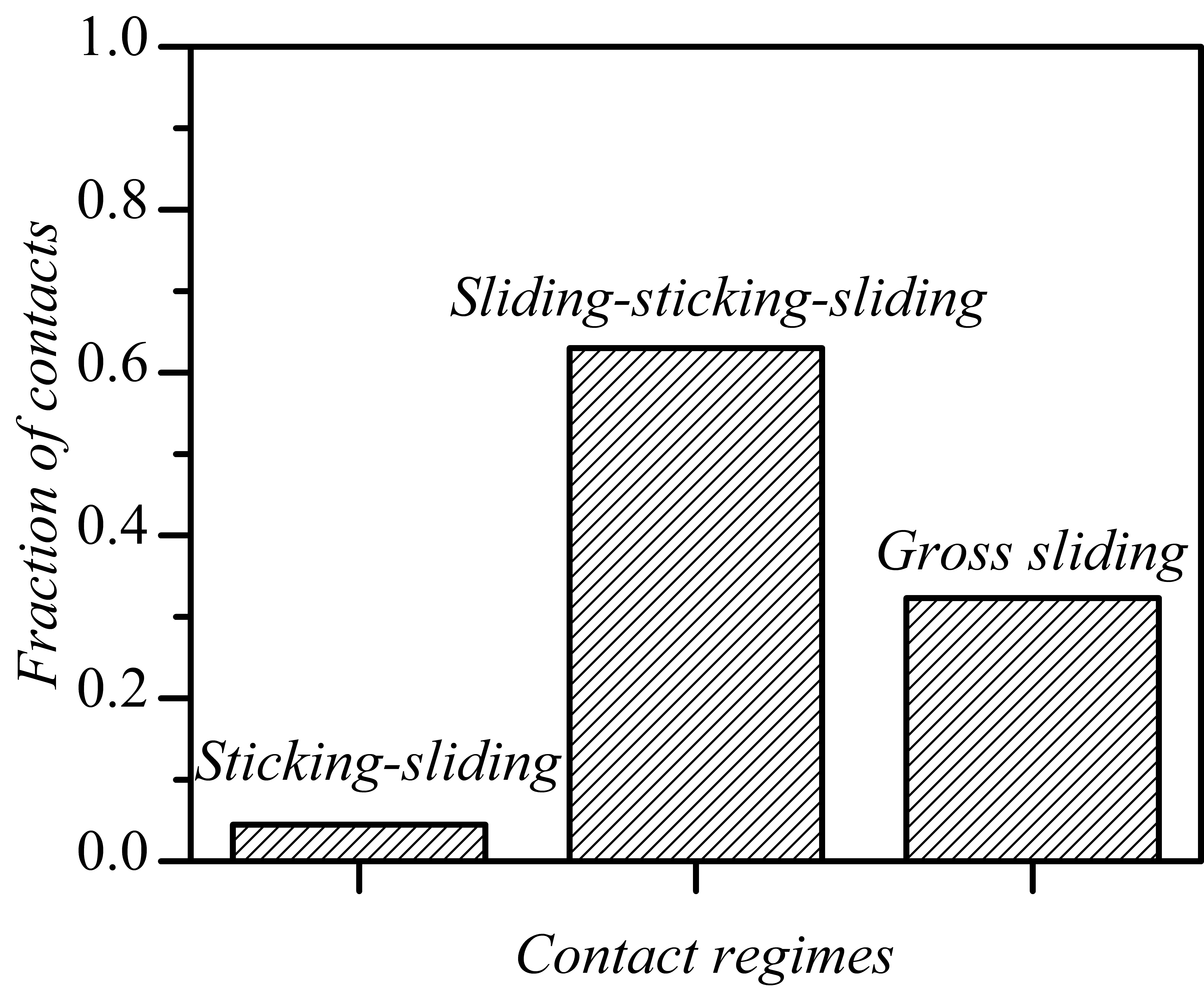}\label{fig:dist_kappa_3_4}}
    \subfloat[c][]{\includegraphics[scale=0.25]{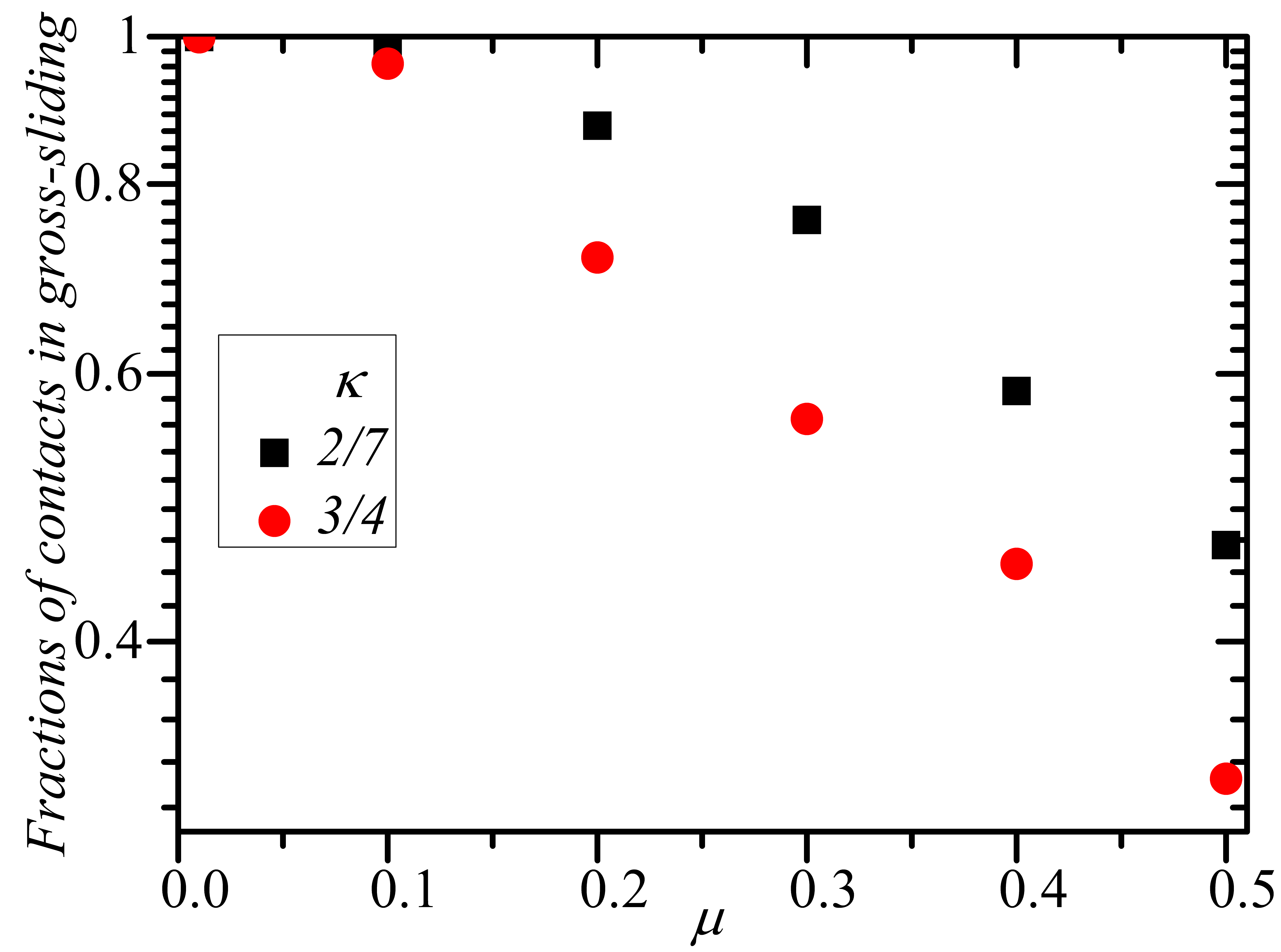}\label{fig:fraction_contacts}}
    \caption{(a) The fraction of the total number of contacts in the sticking and gross sliding regime of $\kappa = 2/7$ (b) Gross sliding, sliding-sticking-sliding, and sticking-sliding regime for $\kappa = 3/4$ in the operating regime at $\mu = 0.5$. (c) The fraction of total number of contacts in the gross-sliding regime for $\kappa = 2/7$ and $3/4$ as the function of inter-particle friction coefficient $(\mu)$ in the log-linear scale.
    }
    \label{fig:distribution_contacts}
\end{figure*}

Next, the fraction of contacts in each of the collision regimes for $\kappa=2/7$ and $3/4$ for different values of inter-particle friction coefficient $(\mu)$ is obtained. Figures \ref{fig:dist_kappa_2_7} and ~\ref{fig:dist_kappa_3_4} show the distribution of contacts in different regimes for $\kappa=2/7$ and $\kappa=3/4$.  The plots suggest the distribution depends on the value of $\kappa$. Fig. \ref{fig:fraction_contacts} shows the fraction of the contact in the gross-sliding regime as a function of $\mu$ for two different values of $\kappa$. It is observed that the percentage of contacts in the gross-sliding regime decreases with the increase in $\mu$. 
We next investigate the influence of $\kappa$ on collective properties of vibrofluidised particles such as linear and angular velocity distribution, volume fraction, granular temperature(s),  and pressure.

\subsection{Effect of $\kappa$ on average kinetic energy and velocity distribution}

\subsubsection{Average kinetic energy} 
\begin{figure*}[]
    \centering
    \subfloat[a][]{\includegraphics[scale=0.25]{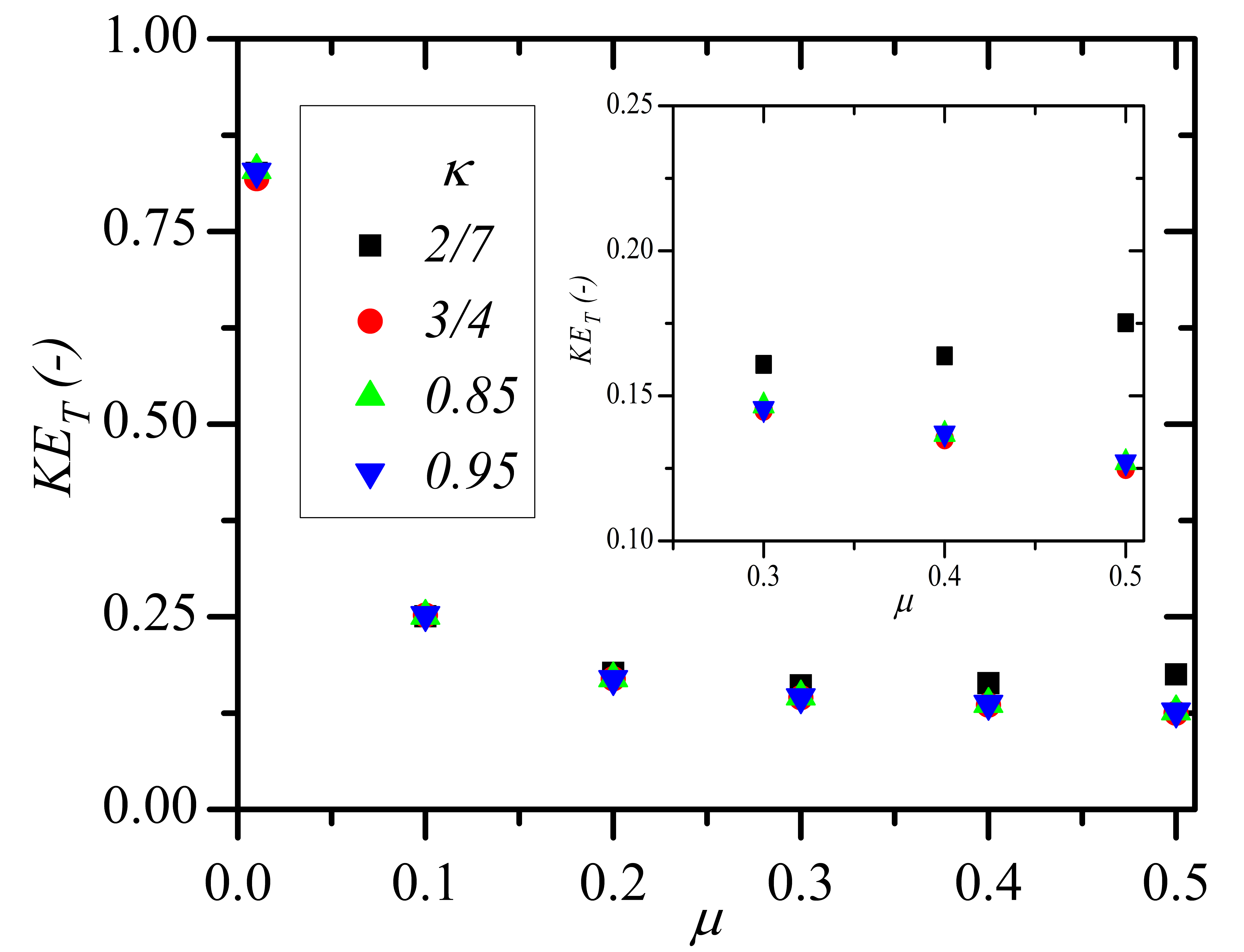}\label{fig:trans_ke_diff_c3}}
    \subfloat[b][]{\includegraphics[scale=0.25]{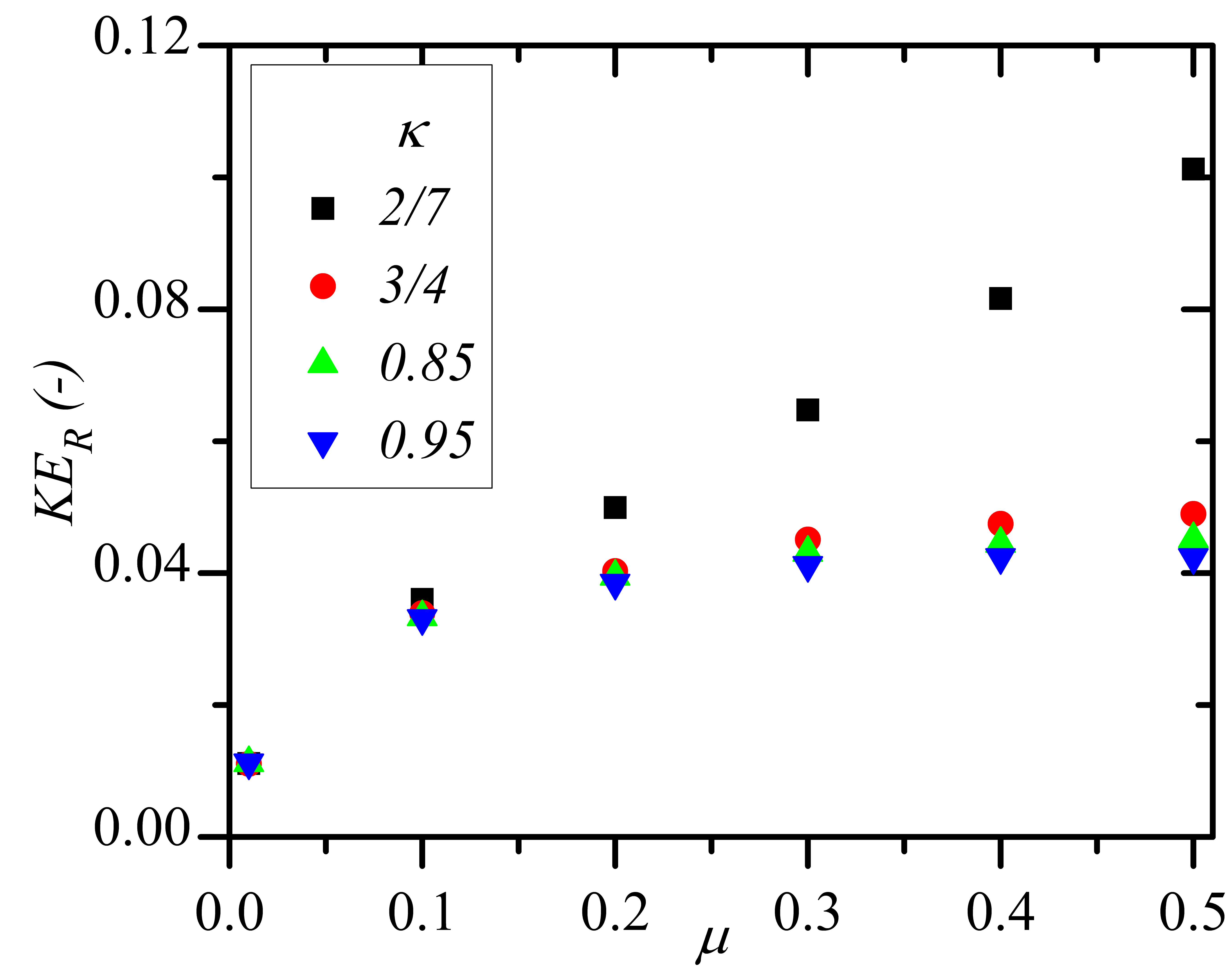}\label{fig:rot_ke_diff_c3}}
     \subfloat[c][]{\includegraphics[scale=0.25]{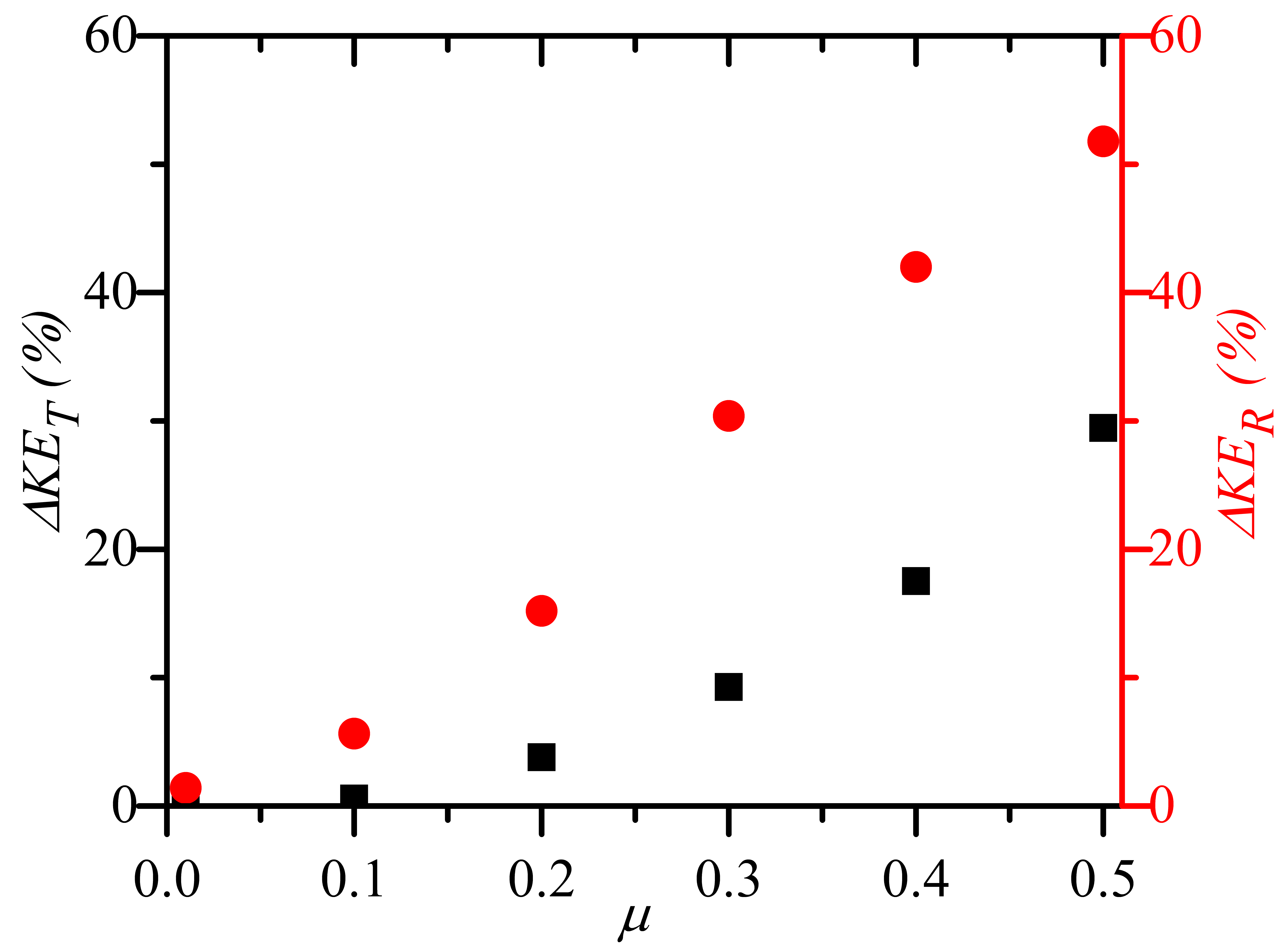}\label{fig:diff_tot_c3}}
         
	\caption{ The plot of (a) Average translational and (b) rotational fluctuating kinetic energy of the particles against the value of inter-particle friction coefficient, (c) $\%\Delta \mathrm{KE}_T$ and $\%\Delta \mathrm{KE}_\Omega$ vs coefficient of friction ($\mu$) for $\kappa= 3/4$. 
    }
    \label{fig:fluc_ke}
\end{figure*}

The mean squared translational $(\mathrm{KE}_T)$ and rotational fluctuating kinetic energy $(\mathrm{KE}_R)$ are plotted against $\mu$ for different values of $\kappa$ (Figure \ref{fig:trans_ke_diff_c3} and \ref{fig:rot_ke_diff_c3}). $\mathrm{KE}_T$ shows a marginal influence of $\kappa$ in the range $0.67 \le \kappa \le 1$. The translational kinetic energy $\mbox{KE}_T$ for $\kappa = 2/7$ is higher than that for other values of $\kappa$, the difference increases with the increase in value of $\mu$. The ratio of tangentian and normal stiffness $\kappa$ has a more prominent influence in $\mathrm{KE}_R$---the higher the value of $\kappa$, lower is $\mathrm{KE}_R$. The \% difference in the values of $\mathrm{KE}_T$ and $\mathrm{KE}_R$ between $\kappa = 3/4$ and $2/7$ for different values of $\mu$ are plotted in Figure \ref{fig:diff_tot_c3}. It is observed that the difference increases with the increase in $\mu$. This difference can be as high as $30\%$ for the $\mathrm{KE}_T$ and 52 \% for $\mathrm{KE}_R$ if $\mu=0.5$. This is because, with the increase in $\mu$, a relatively large fraction of particles follow stick-slip contact rather than gross-sliding (Figure \ref{fig:fraction_contacts}). 

\subsubsection{Velocity distribution}

\begin{figure*}[]
    \subfloat[b][]{\includegraphics[scale=0.25]{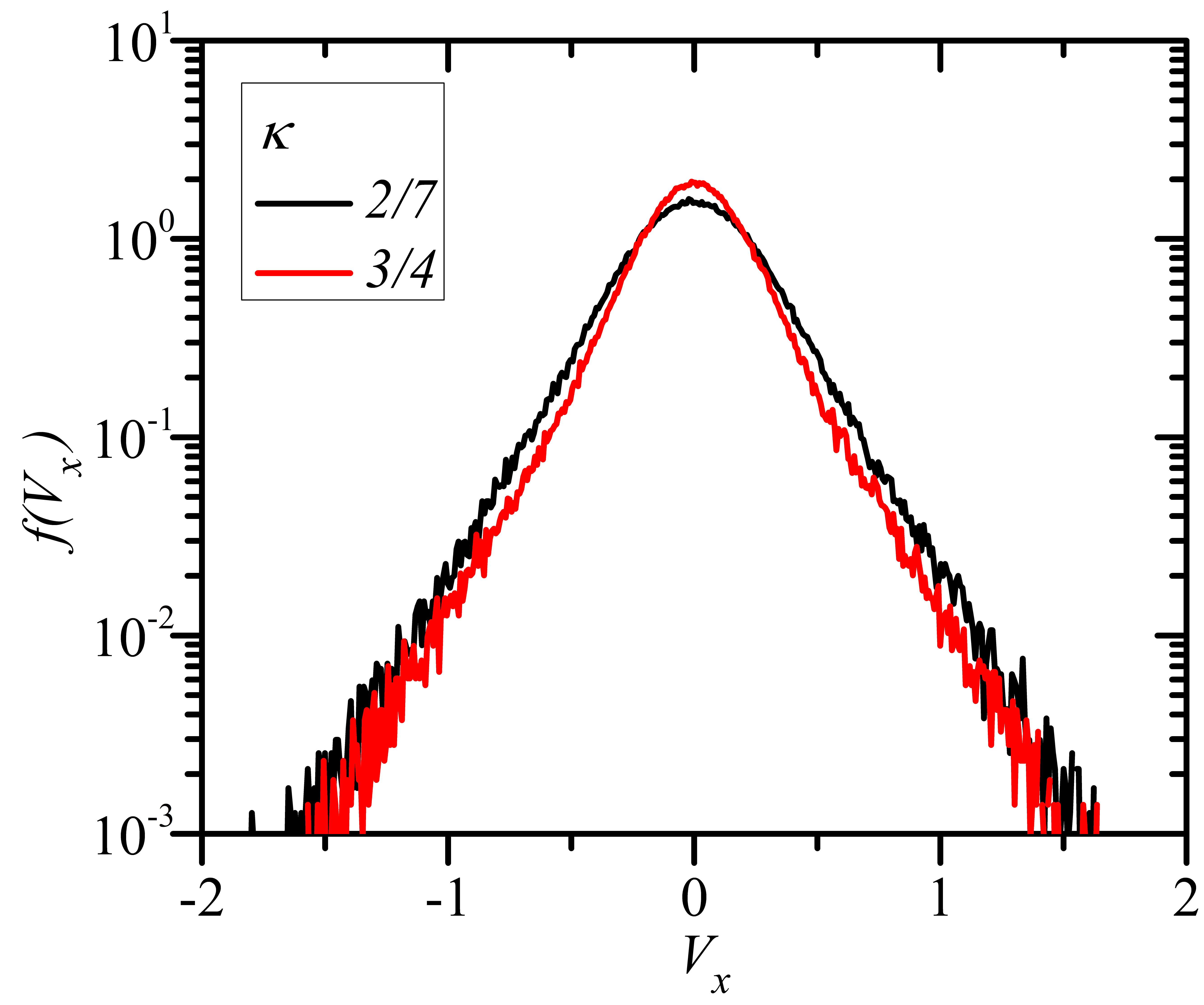}\label{fig:Vx_dis}}
    \subfloat[b][]{\includegraphics[scale=0.25]{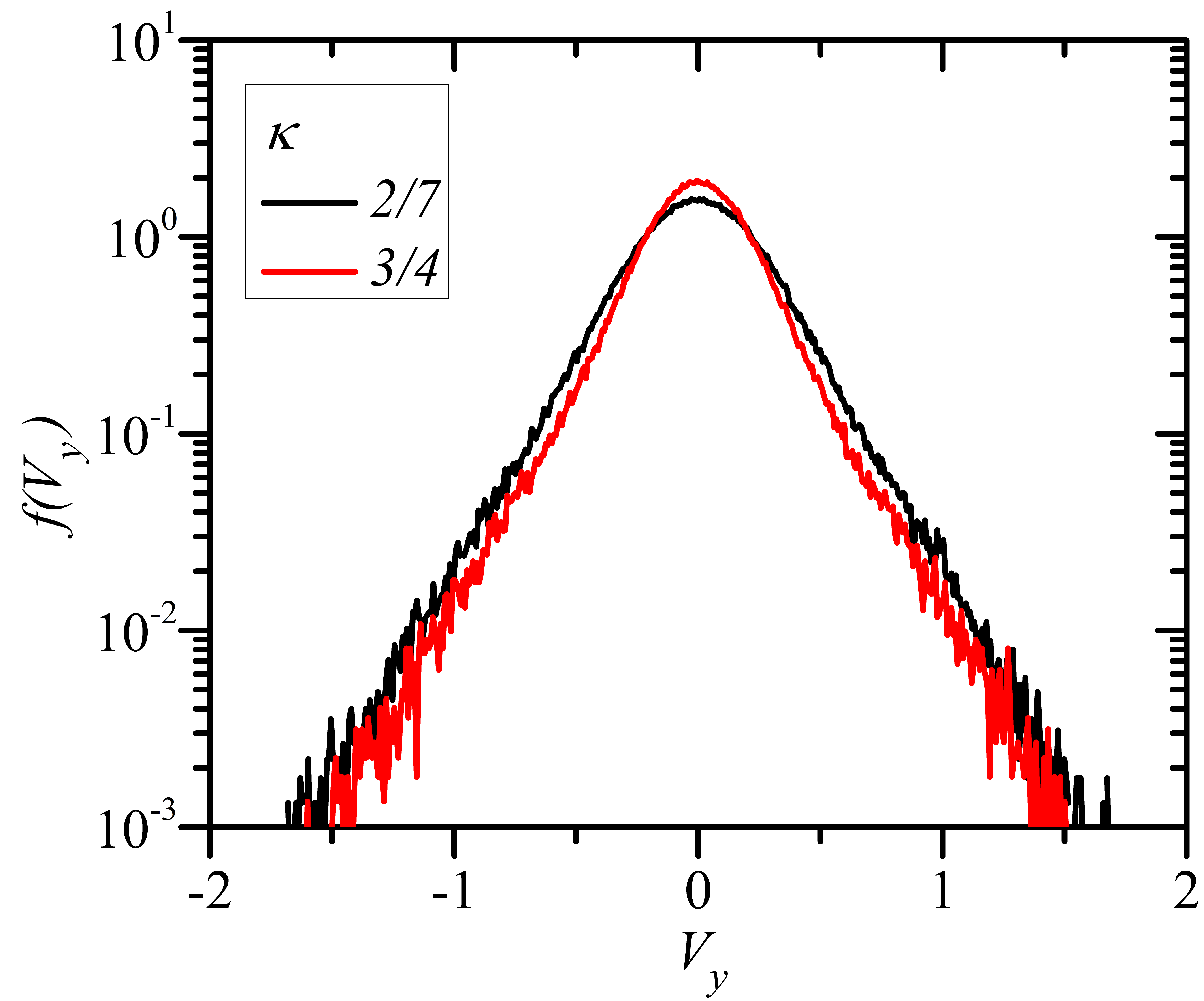}\label{fig:Vy_dis}}
     \subfloat[b][]{\includegraphics[scale=0.25]{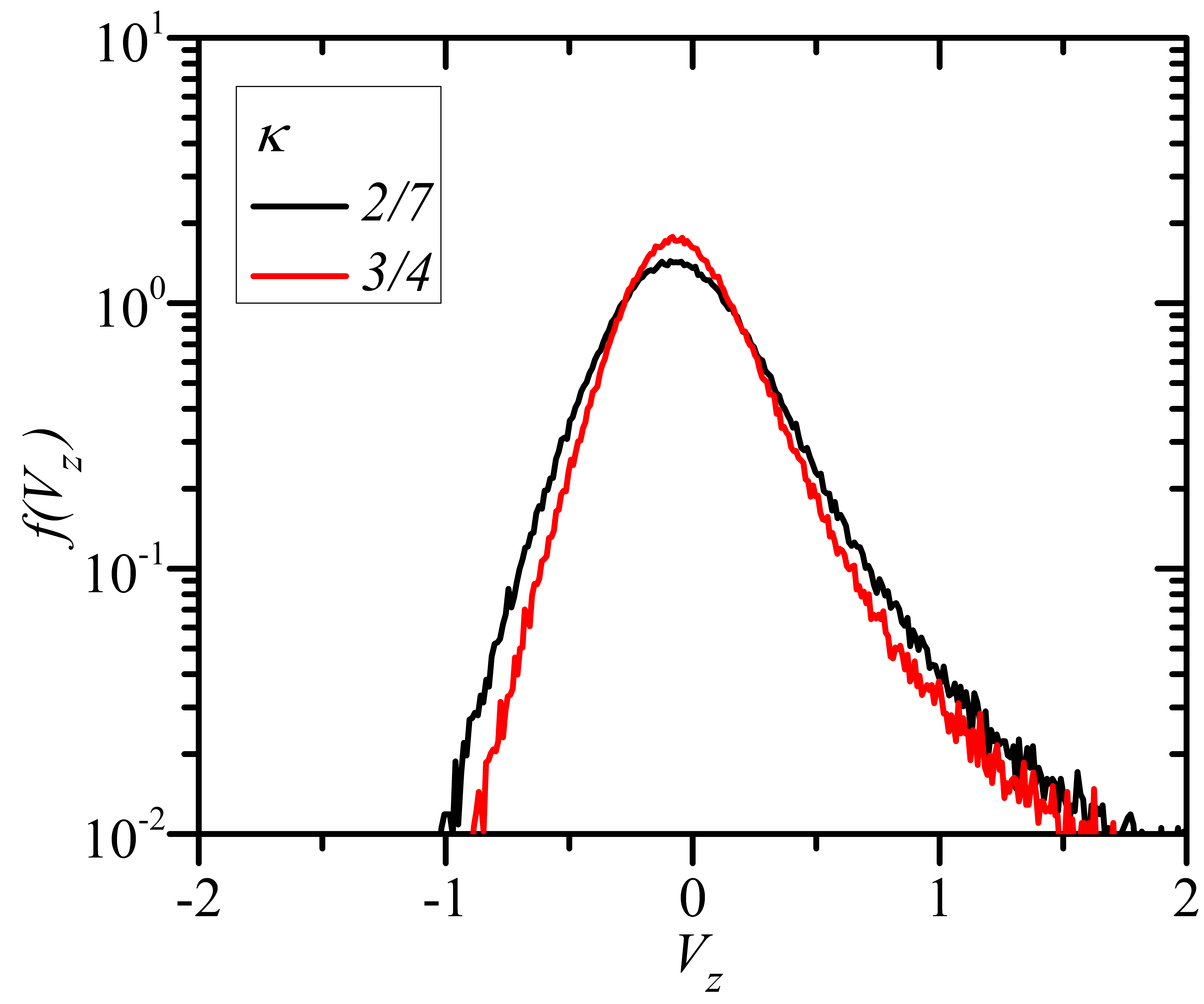}\label{fig:Vz_dis}}

     \subfloat[b][]{\includegraphics[scale=0.25]{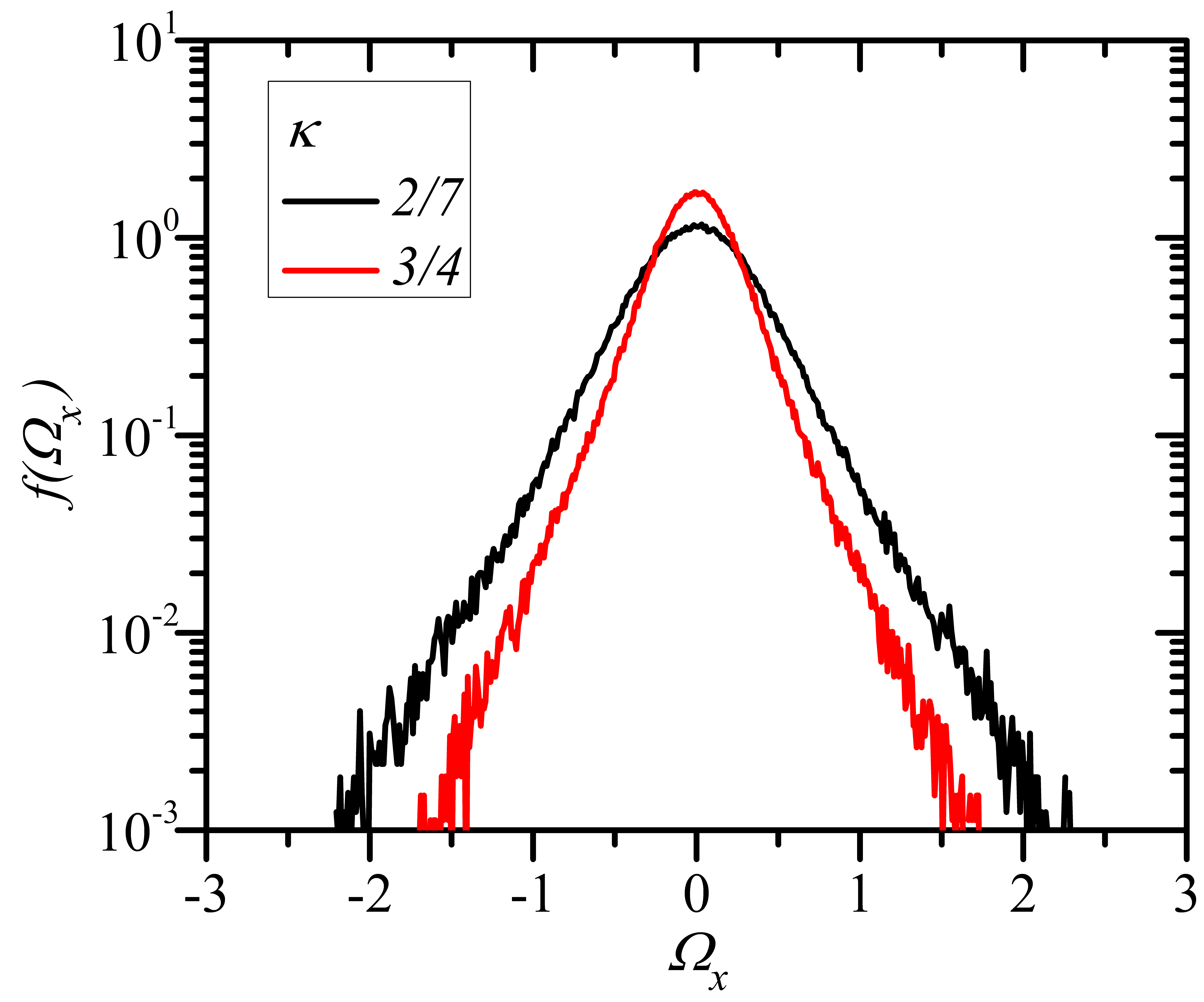}\label{fig:Ox_dis}}   
     \subfloat[b][]{\includegraphics[scale=0.25]{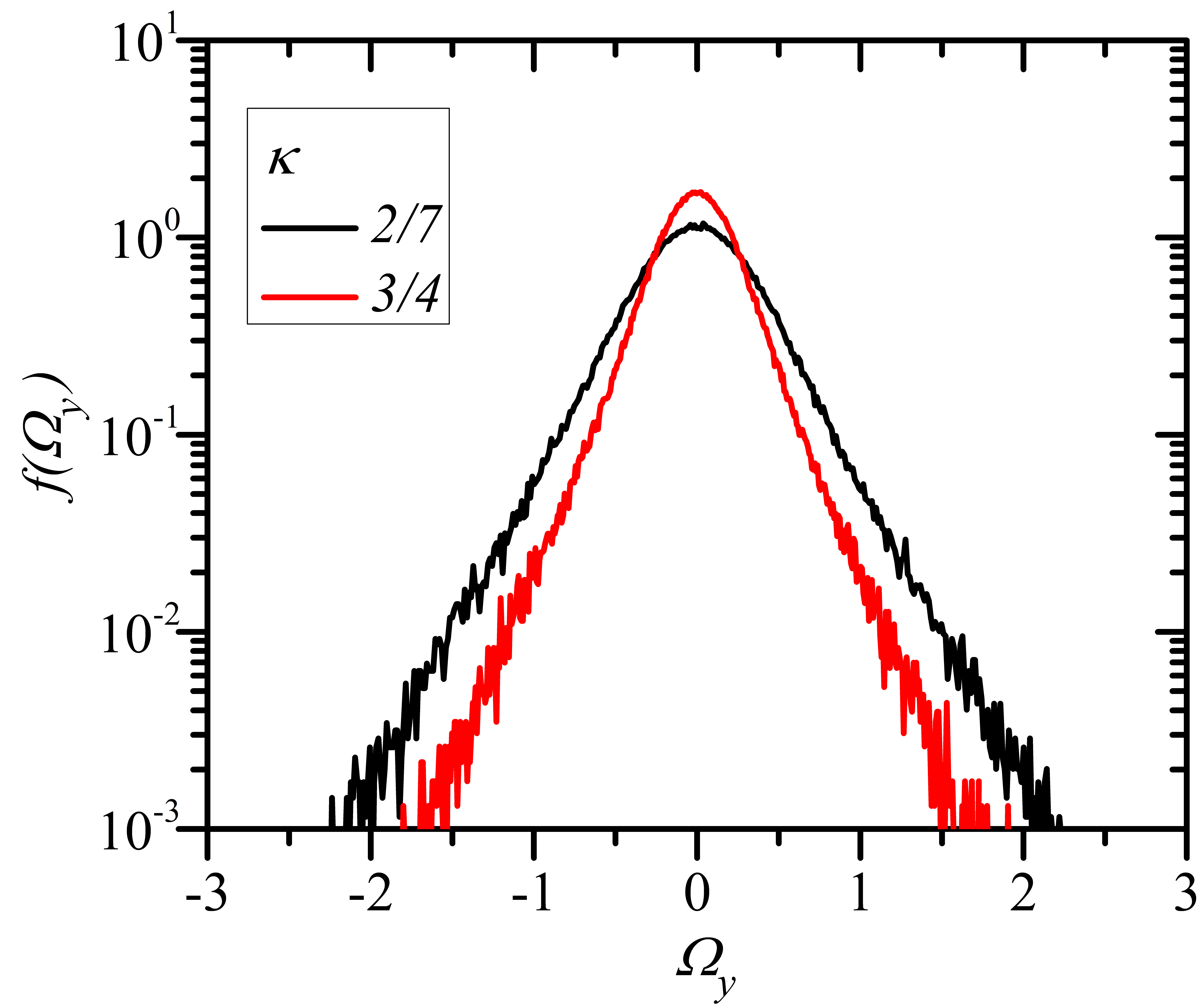}\label{fig:Oy_dis}}
     \subfloat[b][]{\includegraphics[scale=0.25]{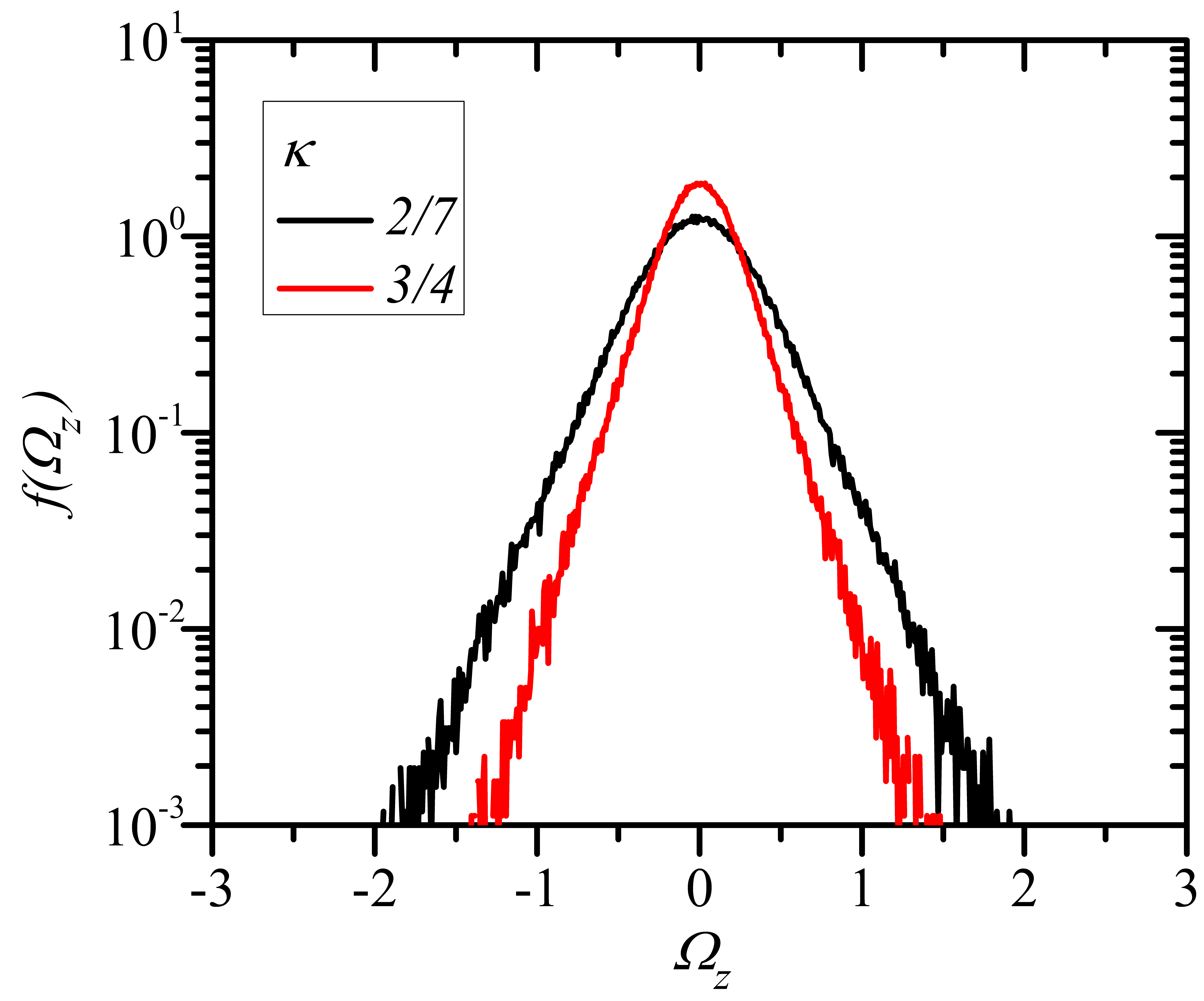}\label{fig:Oz_dis}}

    \caption{Distribution of the components of the linear velocity obtained at steady state for $\mu = 0.5$ and $\kappa=2/7$ and $3/4$,(a) $\mathrm{V_x}$ (b) $\mathrm{V_y}$ (c) $\mathrm{V_z}$ are plotted. Components of the rotational velocities (a) $\Omega_x$ (b) $\Omega_y$ and (c) $\Omega_z$ are plotted.
    }
    \label{fig:linear_rot_vel}
\end{figure*}

 Figure ~\ref{fig:linear_rot_vel} shows the distribution of linear and rotational velocities of particles for two different values of $\kappa (2/7)$ and $(3/4)$ and $\mu=0.5$. The distribution of linear velocity components in the gravity normal direction (x,y in the selected coordinate system) is symmetric. The distribution of the $z$-component of the linear velocity ($V_z$) has a positive skewness of $(N\,u_b^3)^{-1} \Sigma \left(v_z-\bar{v_z}\right)^3 = 2.36$ (Figure \ref{fig:Vz_dis}). The variance for the $v_z$ distribution $(S_z)$ is higher than that for the distribution of $v_x$ and $v_y$, where the variance is defined as $S_{x,y,z} = (N u_b^2)^{-1} \Sigma \left(v_{(x,y,z)} - \bar{v}_{(x,y,z)}\right)^2$. A similar observation is reported in \cite{Krouskop2004,Windows-Yule2013}. The difference between $S_{z}$ and $S_x\,, S_y$ for $\kappa = 2/7$ is $64.5 \%$, whereas for $\kappa=3/4$ it is $71.6 \%$ for the case with $\mu=0.5$. The difference decreases with the decrease in value of $\mu$. A similar exercise is carried out for the rotational velocity of the particles. The variance of the distribution of the z-component of the angular velocity ($S^{\Omega}_{x,yz} = (d^2/N u_b^2) \Sigma \left(\omega_{(x,y,z)} - \bar{\omega}_{(x,y,z)}\right)^2)$ is about $22\%$ smaller than $S^{\Omega}_x, S^{\Omega}_y$ irrespective of the values of $\kappa$.

\subsection{Effect of $\kappa$ on the profile of solid volume fraction, granular temperature and pressure}

\begin{figure*}[]
    \centering
    \subfloat[a][]{\includegraphics[scale=0.25]{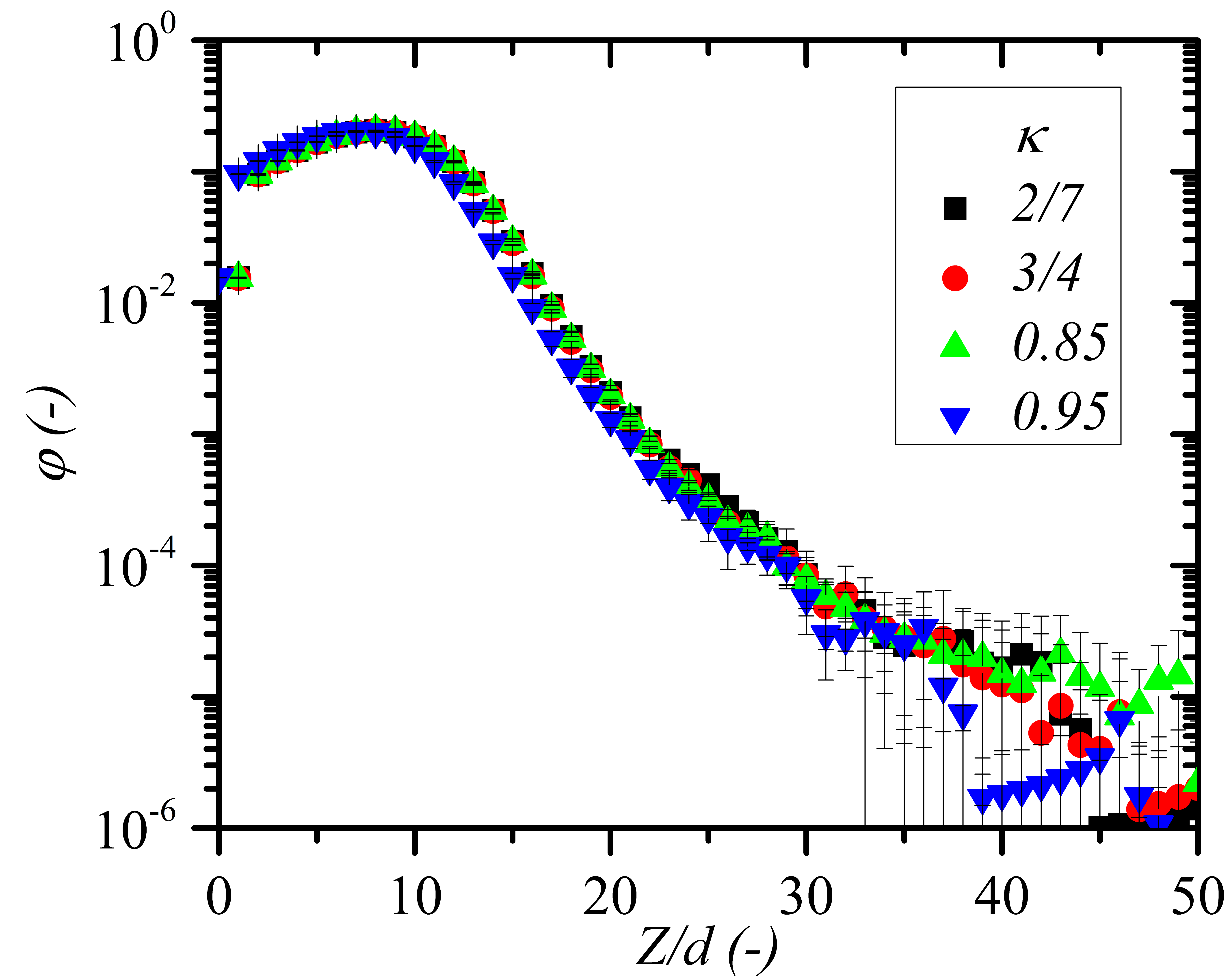}\label{fig:vol_fraction_0.1}}
    \subfloat[b][]{\includegraphics[scale=0.25]{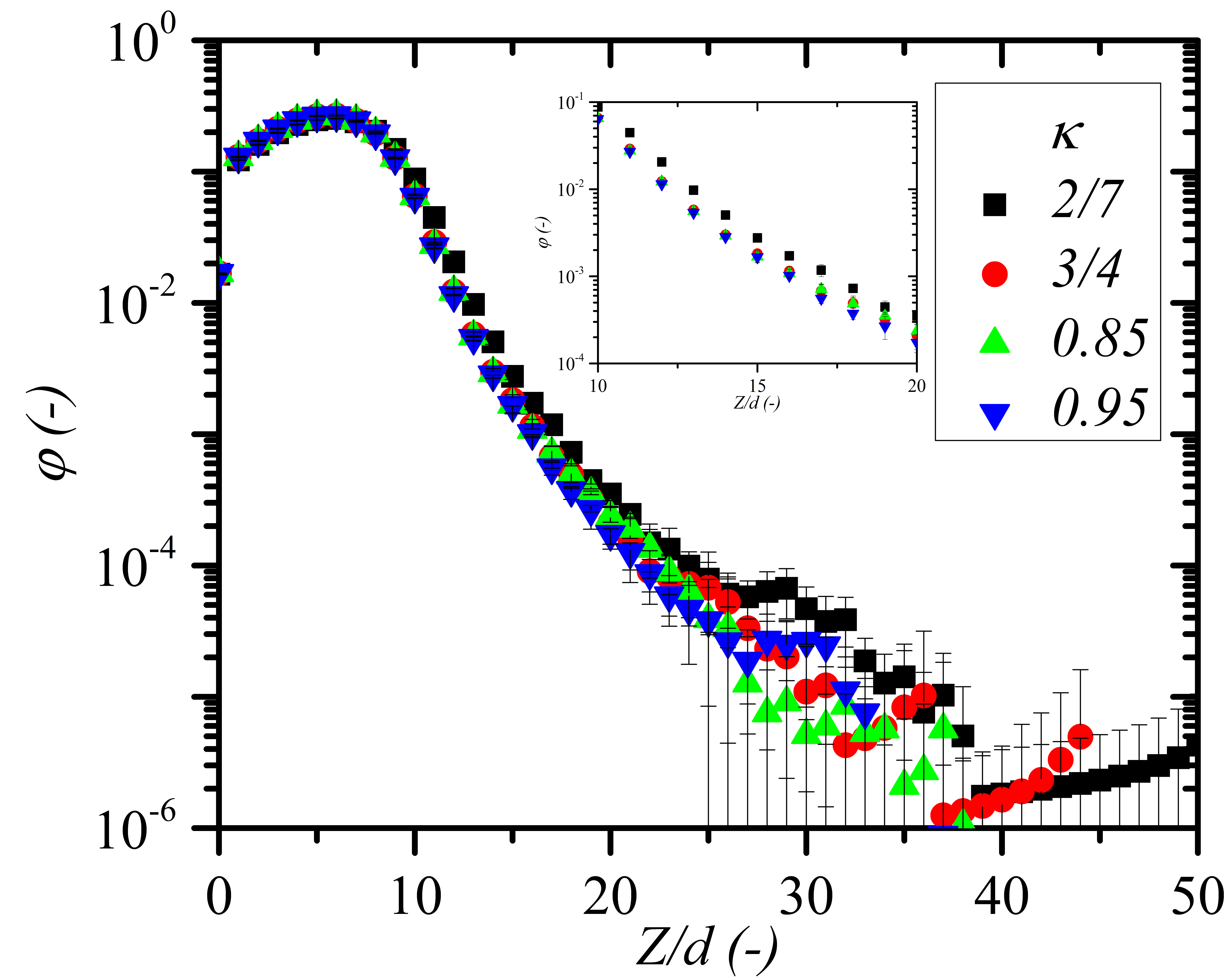}\label{fig:vol_fraction_0.3}}
     \subfloat[c][]{\includegraphics[scale=0.25]{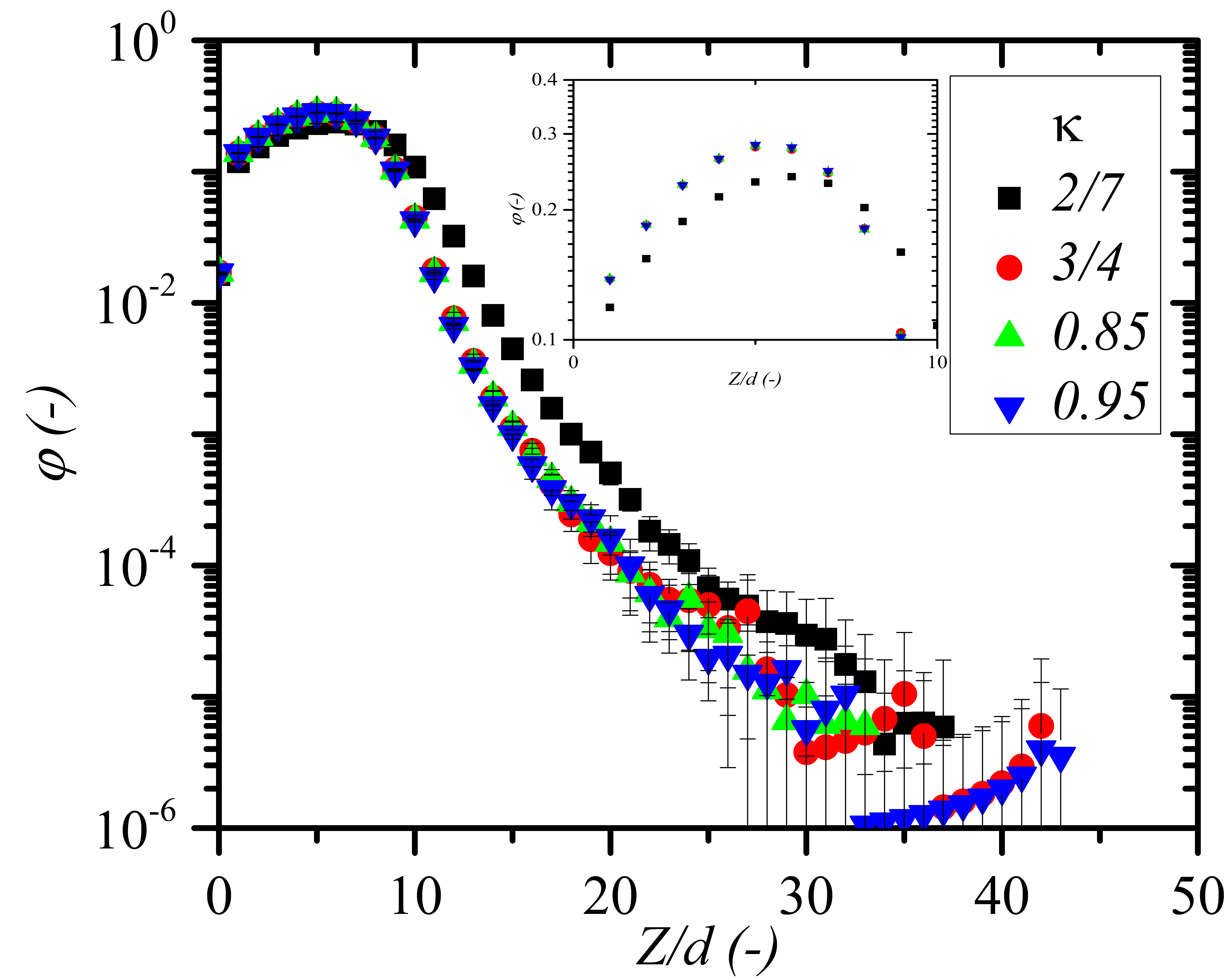}\label{fig:vol_fraction_0.5}}\\
     \subfloat[b][]{\includegraphics[scale=0.25]{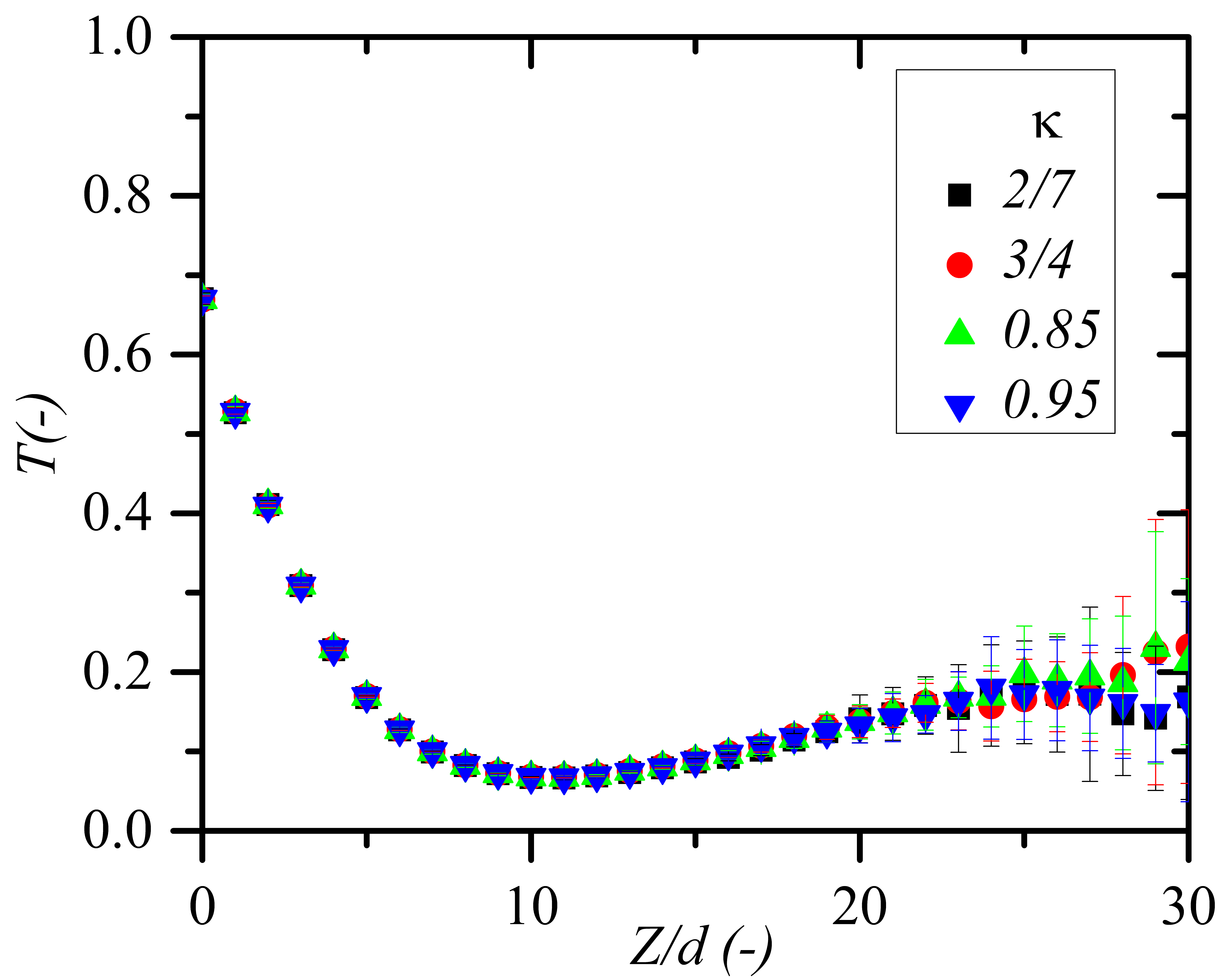}\label{fig:trans_mu_0.1}}   
     \subfloat[b][]{\includegraphics[scale=0.25]{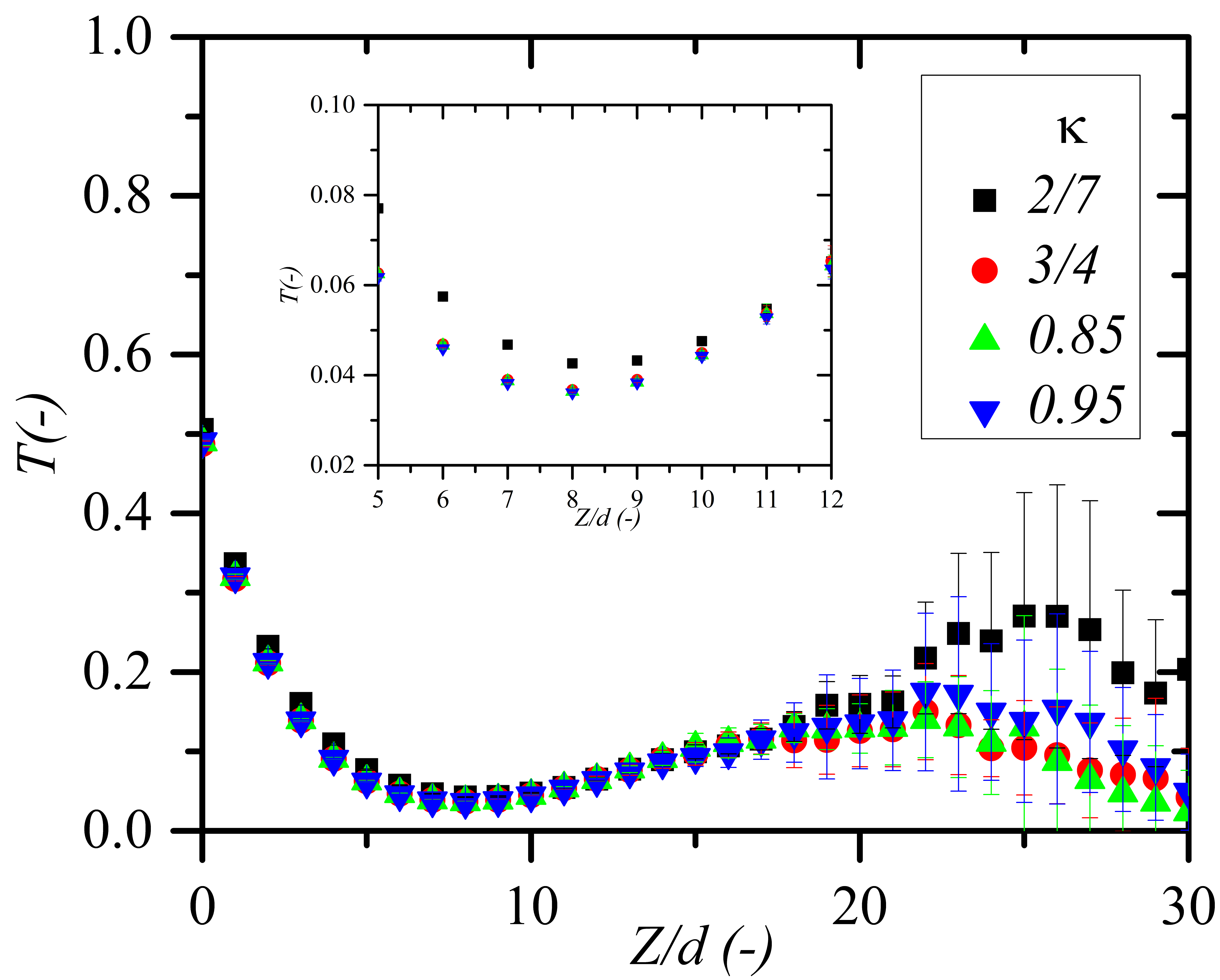}\label{fig:trans_mu_0.3}}
     \subfloat[b][]{\includegraphics[scale=0.25]{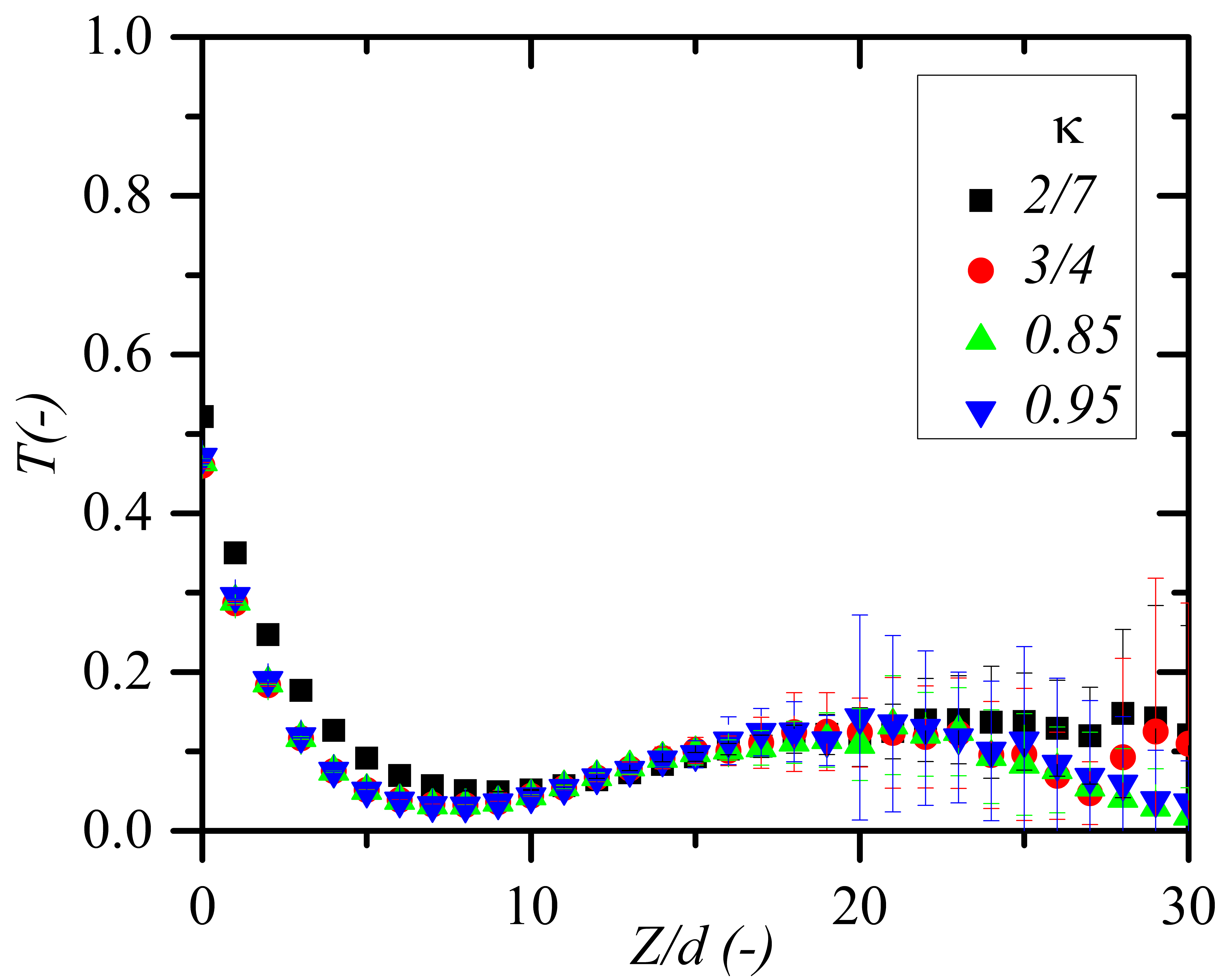}\label{fig:trans_mu_0.5}}
       \\
      \subfloat[b][]{\includegraphics[scale=0.25]{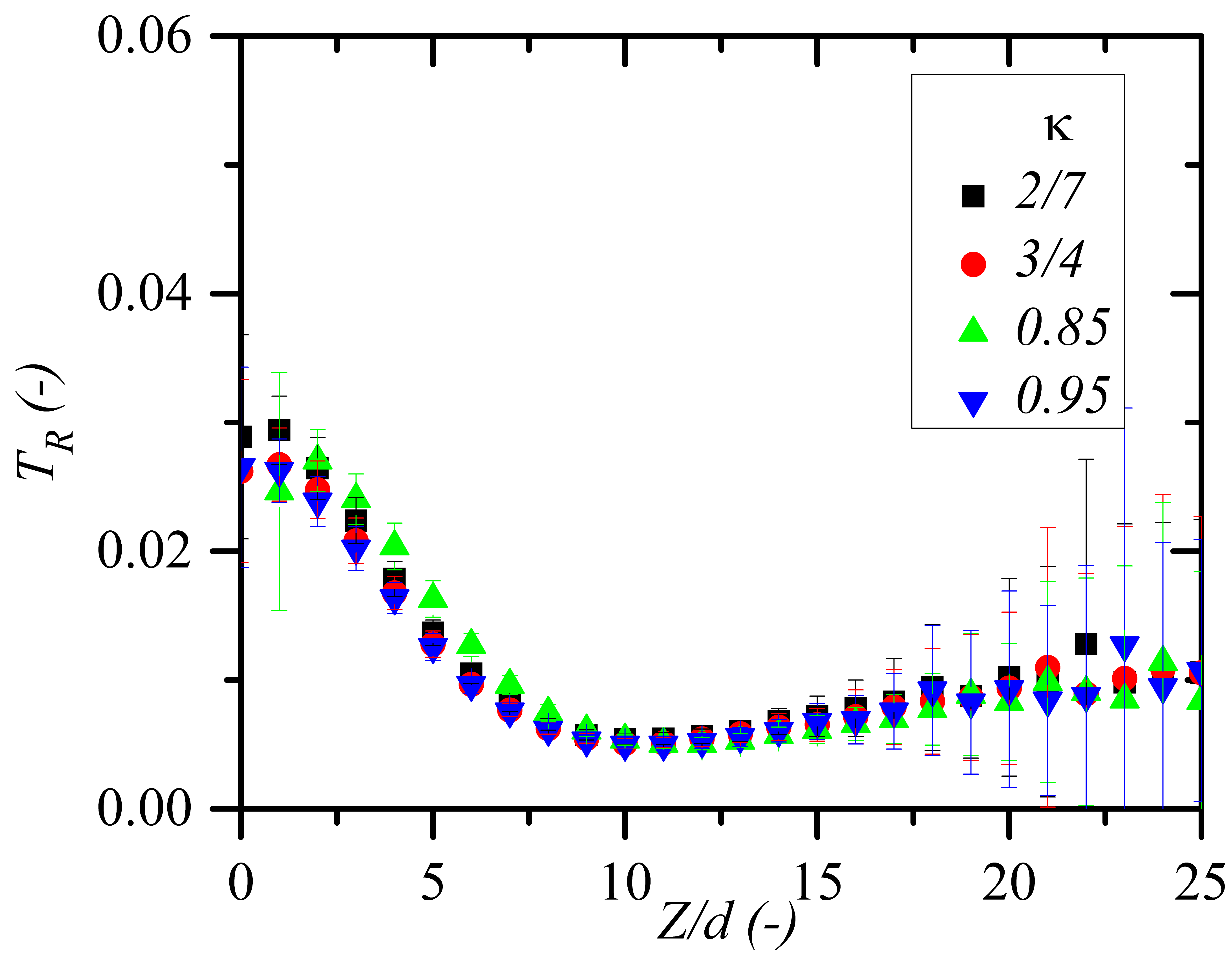}\label{fig:rot_mu_0.1}}   
     \subfloat[b][]{\includegraphics[scale=0.25]{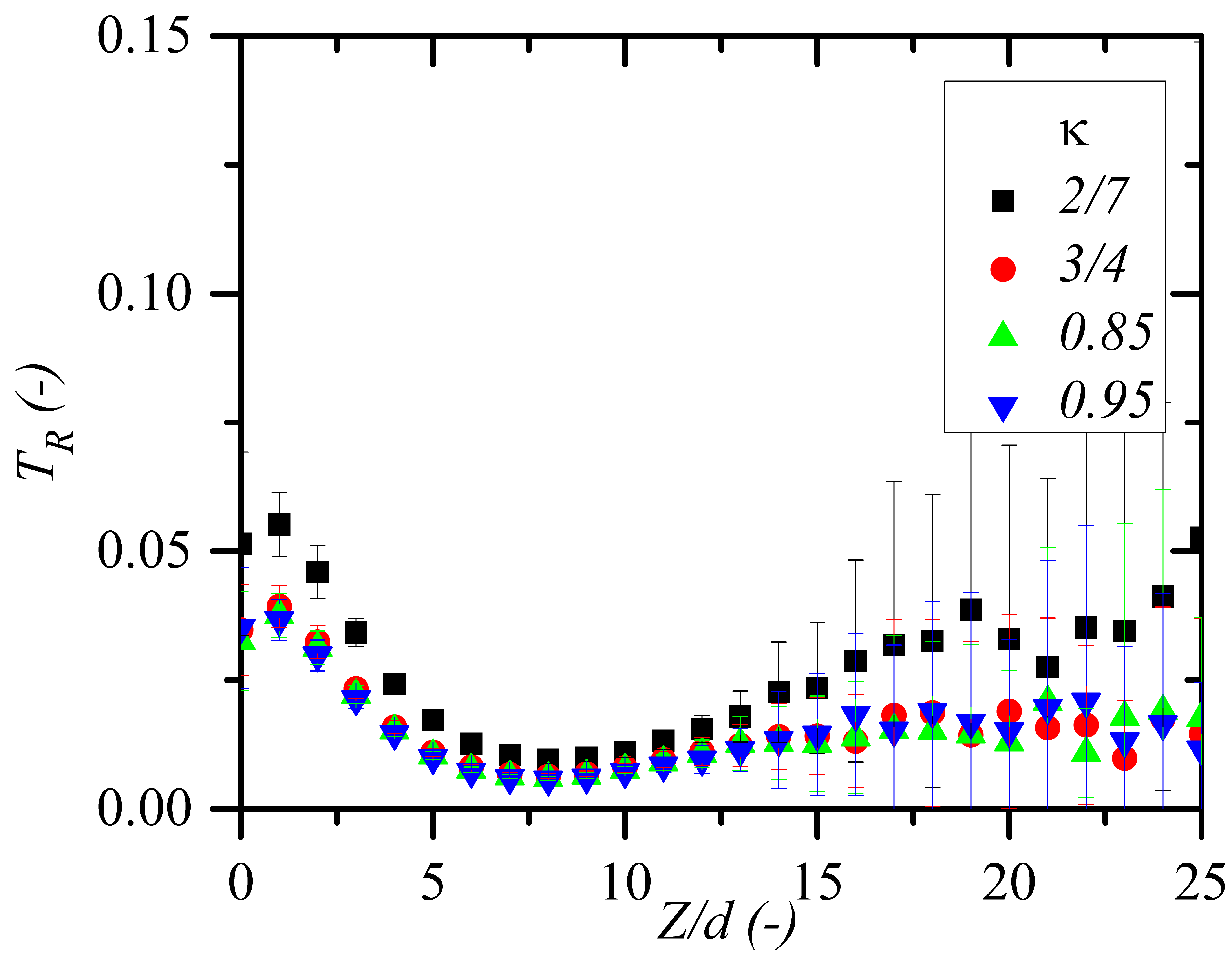}\label{fig:rot_mu_0.3}}
     \subfloat[b][]{\includegraphics[scale=0.25]{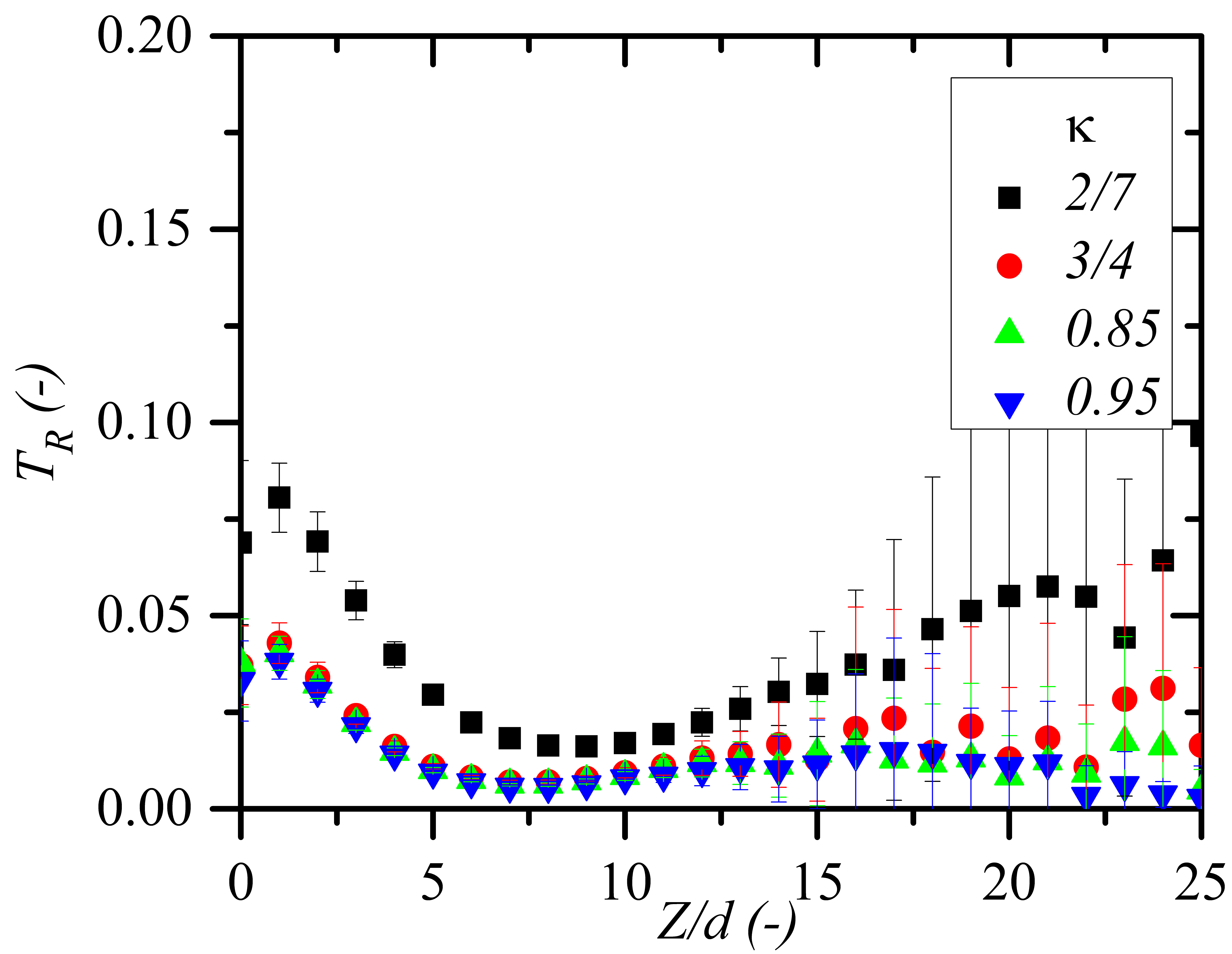}\label{fig:rot_mu_0.5}}\\
      \subfloat[b][]{\includegraphics[scale=0.25]{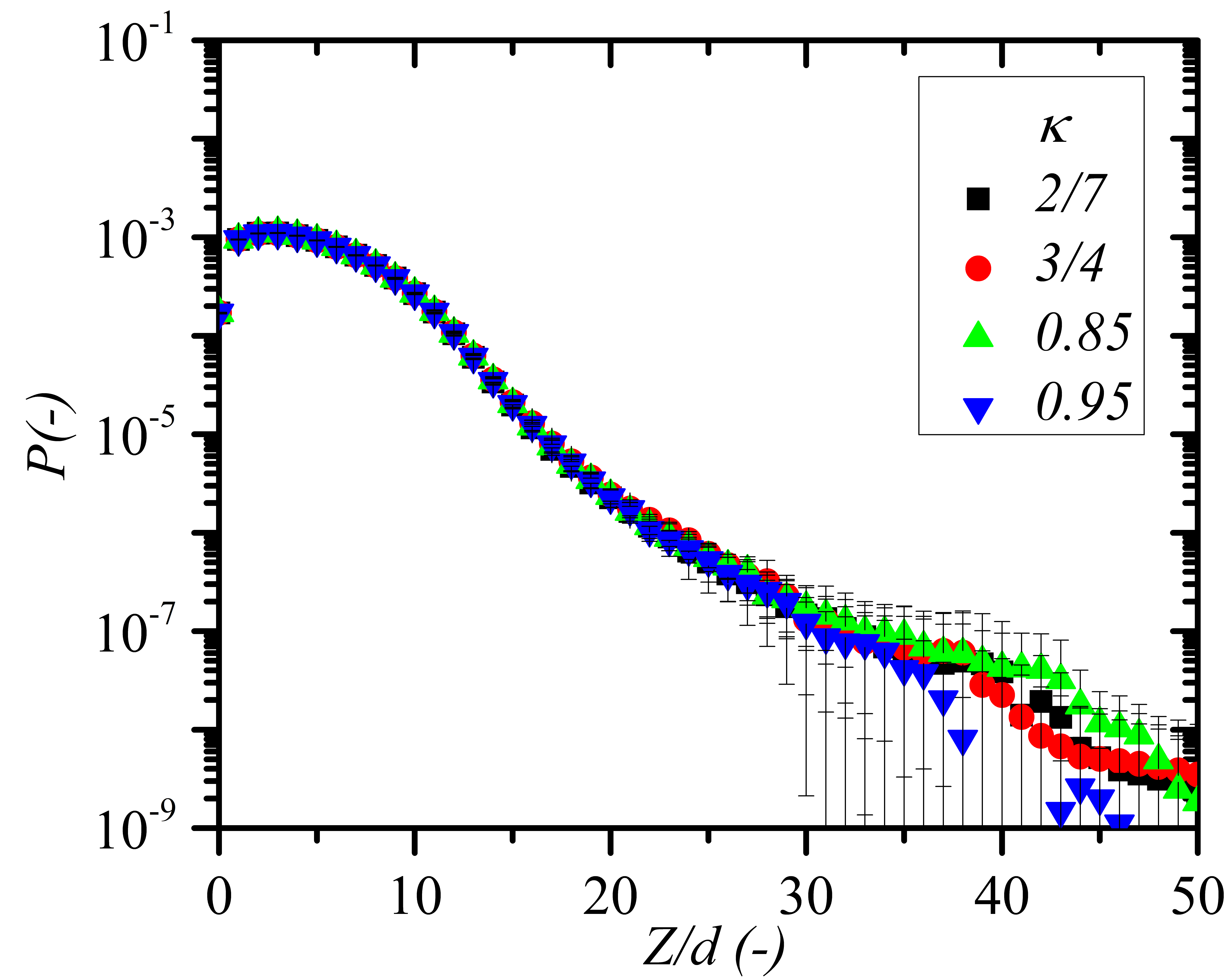}\label{fig:tot_stress_0.1}}   
     \subfloat[b][]{\includegraphics[scale=0.25]{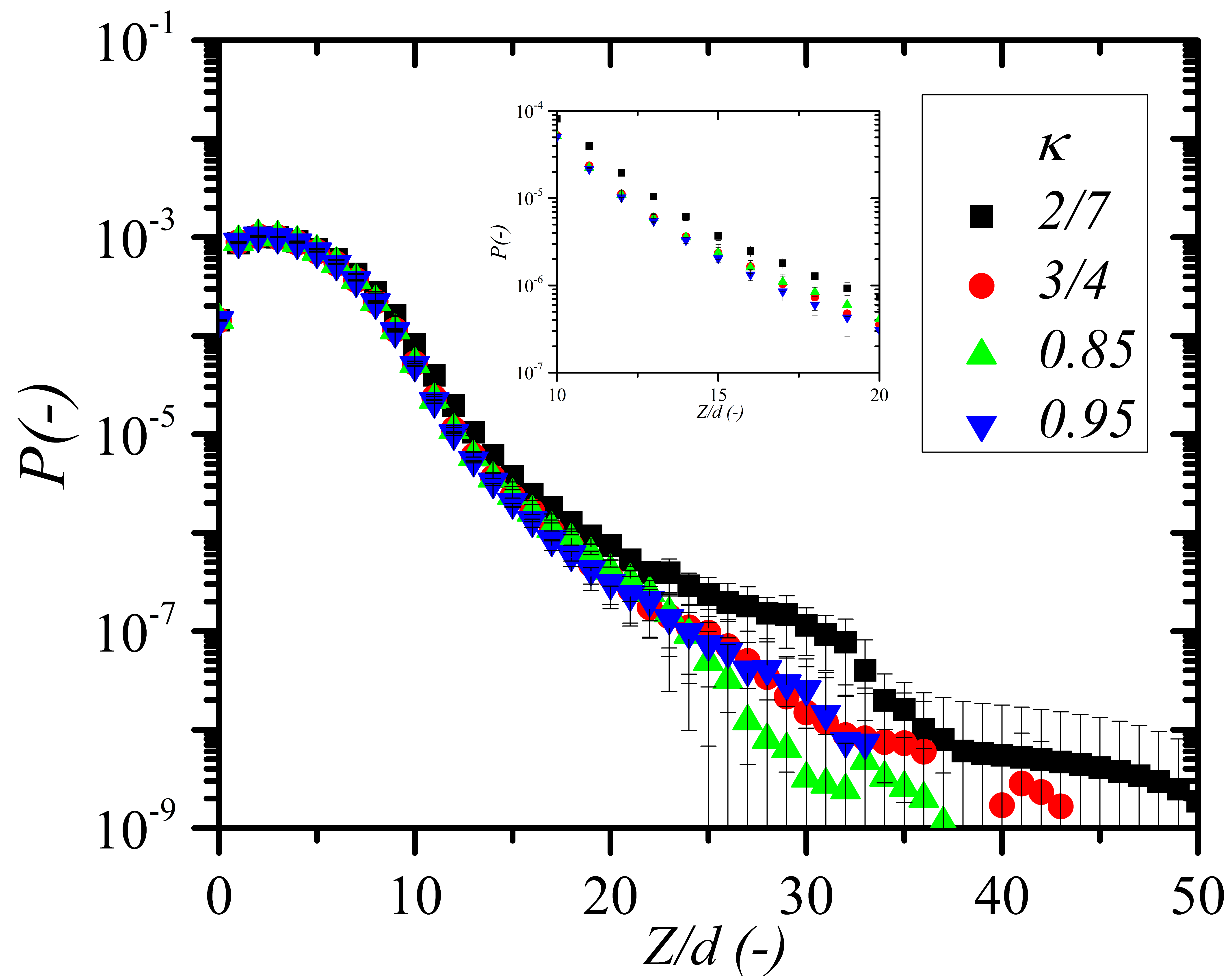}\label{fig:tot_stress_0.3}}
     \subfloat[b][]{\includegraphics[scale=0.25]{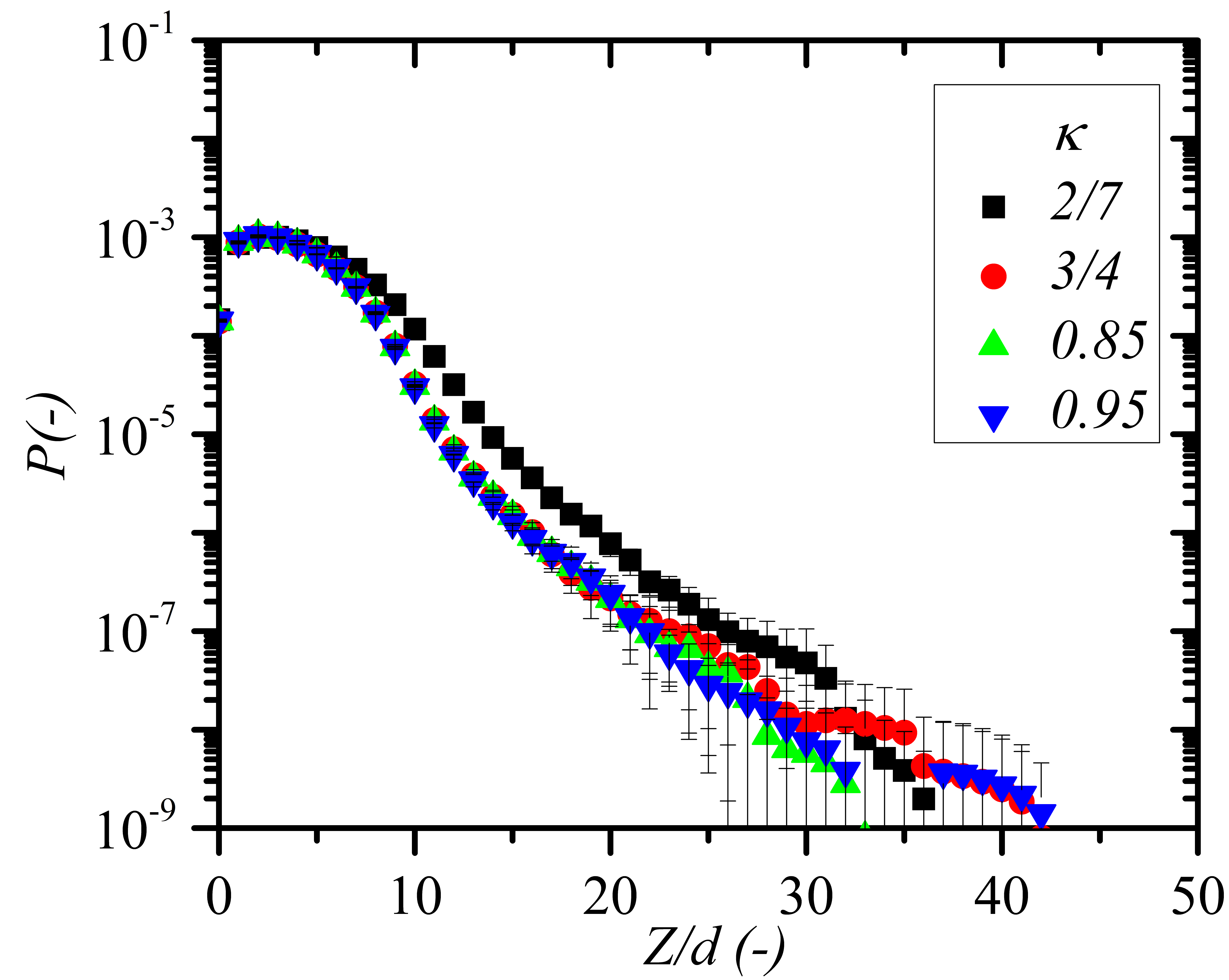}\label{fig:tot_stress_0.5}}

    \caption{Solid volume fraction (a) $\mu$ =0.1 (b) $\mu$ =0.3 and (c) $\mu$ = 0.5; translational granular temperature ($T$) (d) $\mu$ =0.1, (e) $\mu$ = 0.3 and (f) $\mu=0.5$; Rotational granular temperature $(T_R)$ (g) $\mu$ =0.1, (h) $\mu$ = 0.3 and (i) $\mu=0.5$;
    Pressure for (j) $\mu$ = 0.1, (k) $\mu$ = 0.3 and (l) $\mu$ = 0.5}
    \label{fig:vol_fraction}
\end{figure*}

Profiles of the macroscopic properties, such as the solid volume fraction, granular temperature, and pressure, for different values of $\kappa$ and $\mu$, are plotted in Figure \ref{fig:vol_fraction}. The profiles of solid volume fraction plotted in Figures ~\ref{fig:vol_fraction_0.1} to ~\ref{fig:vol_fraction_0.5} exhibit a non-monotonic behaviour as reported for vibrofluidised particles in a vertical bed in earlier literature \citep{Brey2001}. 
 
Figures ~\ref{fig:trans_mu_0.1} to ~\ref{fig:rot_mu_0.5} show the 
translational and rotational granular temperatures for four different values of $\kappa$ and $\mu = 0.1,0.3$ and $0.5$. Both translational and rotational granular temperature exhibit minima at an intermediate height. Similar behaviour is reported in \citep{Helal1997}, who argued that the minimum in the translational granular temperature is due to the dependence of the conductivity of the granular temperature on the solid volume fraction. The ratio of tangential and normal stiffness $\kappa$ has a marginal influence on the translational granular temperature for $\mu = 0.1$; however, for $\mu = 0.3$ the translational granular temperature near the base for $\kappa=2/7$ is nearly 20\% higher than the others. The effect of larger temperature is reflected in the slower decay in the volume fraction profile (inset of Figure 9b). The effect of $\kappa$ on the rotational granular temperature for $\mu = 0.3$ and above is clearly visible from the figure.

Figures \ref{fig:tot_stress_0.1} to ~\ref{fig:tot_stress_0.5} plot the profiles of the pressure in a vertically vibrated granular bed. Again, the effect of $\kappa$ on the pressure is observed only if $\mu = 0.3$; for $\mu = 0.1$, $\kappa$ has no influence on pressure.

\section{Conclusion}

A granular assembly in a vertical vibrofluidised bed is simulated using the linear-spring dashpot model on the open-source platform LAMMPS for different values of the ratio of the tangential to normal spring stiffness $\kappa=k_t/k_n$ and the inter-particle coefficient of friction $(\mu)$. The value of the normal spring stiffness is selected to be $10^8 mg/d$ so that the contacts are mainly binary. The rotational coefficient of restitution is determined using the pre- and post-collision linear and angular velocities of the particles at contact. The following are the conclusions from this study.

\begin{enumerate}
    \item There exist two mutually exclusive sticking and sliding regimes for $\kappa=2/7$; however, for $\kappa$ in the range of 0.67 to 1, sliding-sticking-sliding and gross sliding regimes are observed.

    \item Near-normal contact behaviour is captured using a linear-spring dashpot model; the transition between the near-normal to the intermediate regime is marginally different from that suggested by \cite{Maw1976}.

    \item For the inter-particle coefficient of friction more than 0.1, a significant portion of contacts are in the non-gross sliding regime for all the values of $\kappa$.

    \item The mean-squared translational and rotational fluctuating kinetic energy of the particles in the vibrated bed significantly differ with the choice of $\kappa$ for the inter-particle coefficient of friction more than 0.1.

    \item Profiles of granular temperature (mean-fluctuating kinetic energy), solid volume fraction and pressure are plotted and are shown to differ with the choice of $\kappa$ for $\mu > 0.1$.

\end{enumerate}

To summarise, the current work shows the effect of tangential spring stiffness constant on the linear and angular velocity distributions, profiles of the solid volume fraction, granular temperature, and pressure of the vibrofluidised particles, all of which are found to be quantitatively different for $\kappa = (2/7)$ in comparison to those in the range $0.67\leq \kappa < 1 $  and $\mu > 0.1$. It is interesting to investigate the effect on systems like flow over an inclined chute or sheared granular material.

\begin{acknowledgments}
We thank the Central Library of the Indian Institute of Technology for the licensed version of Grammarly. The software was used for the English Grammar check of the manuscript. VK was supported by funding from the MHRD and the Science and Engineering Research Board, Government of India (Grant no. SR/S2/JCB-31/2006).
\end{acknowledgments}

\appendix
\counterwithin{figure}{section}

\section{Algorithm for pair detection}
\label{appendix:algo}
\begin{figure}[]
\centering
\includegraphics[width=7cm,height=18cm,keepaspectratio]{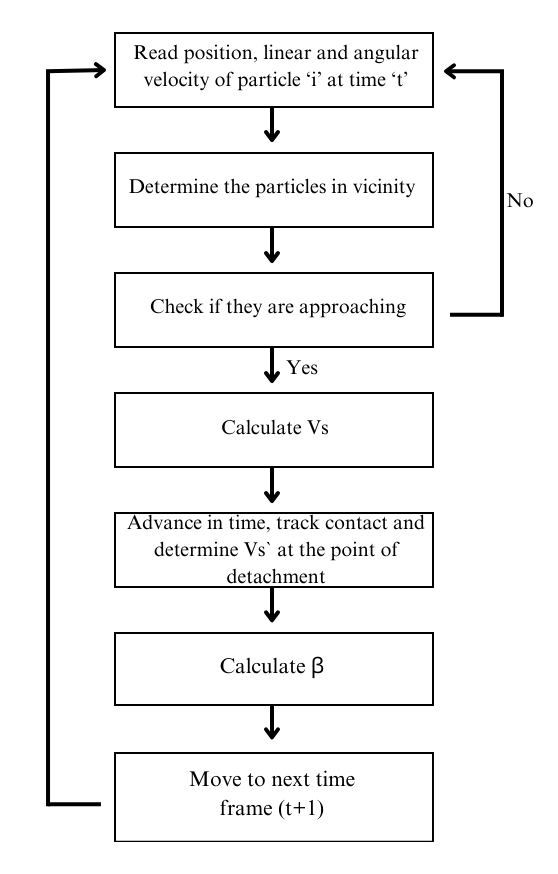}
\caption{Flow chart for determining pre and post-collisional velocities, rotational coefficient of restitution $(\beta)$ from the DEM simulation data.  }\label{fig:algorithm}
\end{figure}

The flow-chart (Figure ~\ref{fig:algorithm}) presents the algorithm to determine the pair of particles at contact from DEM data. At any time `t', particles in the neighbour of any particle `i' are obtained such that inter-particle distance is $ d \leq \left \| \vec{r}_{ij} \right \| \leq d + \epsilon$, where $\epsilon$ is $1\%$ of the diameter of the particle. Particle `j' from the list of neighbours is said to approach or collide with particle `i' when $\left ( \vec{v}_{ij} \cdot \hat{n}_{ij} \right ) < 0$. Once a colliding pair is identified at the pre-collision time `t', the collision is tracked further in time by advancing simulation with the time-step of $\delta t$. The post-collision properties are obtained at the time-step where $\left ( \vec{v}_{ij} \cdot \hat{n}_{ij} \right ) > 0$ and $ \left \| \vec{r}_{ij} \right \| > d$ are satisfied. Once pre- and post-collision time-step of the interacting particles are identified, simulation data are processed further for the following.

\begin{enumerate}

\item To determine the coordination number (CN) defined as the average number of particles at contact with another at any instance of time \citep{Reddy2010};
\item To determine the frequency distribution of CN normalized with the total number of samples.

\item To determine the rotational coefficient of restitution $(\beta = - v^{*}_{s}/ v_{s})$.

\item To obtain the non-dimensionalized impact and rebound angle, $\psi_1 =  (\kappa/\mu) (v_s/v_n)$ and $ \psi_{2} =  (\kappa/\mu) (v^{*}_s/v_{n})$, respectively.

\item To determine the evolution of normal and tangential force during the contact.

\end{enumerate}

\section{Time-step convergence}
\label{appendix:time_convergence}

\begin{figure}[]
\centering
\includegraphics[scale = 0.25]{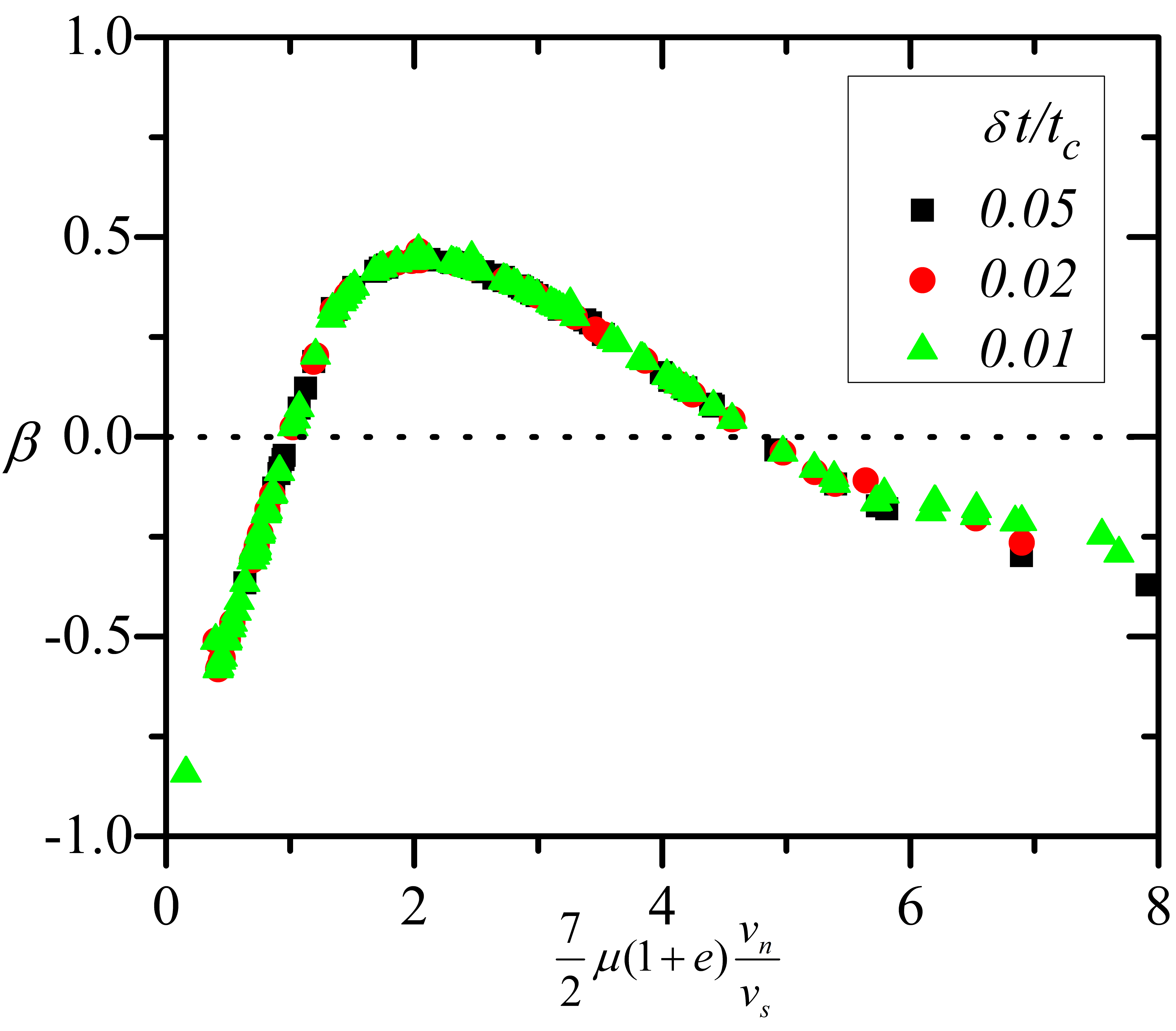}
\caption{ Plot of rotational coefficient of restitution $(\beta)$ against $(7/2) \mu (1+e_n) (v_n/v_s)$ for varying simulation time-step $(\delta t)$. }\label{fig:time-step_conv}
\end{figure}

The DEM time-step $(\delta t)$ is varied from ${1/20}^{th}$ to ${1/100}^{th}$ of the collision time. It is observed that the scaling of $\beta$ against $v_{n}/v_{s}$ does not vary with the selection of DEM time-step if contact is resolved by at least ${1/20}^{th}$ of the collision time-step $t_c$ (Fig. \ref{fig:time-step_conv}). 

\section{Discussion on $\beta$}
\label{derivation}
The rotational coefficient of restitution, defined as the ratio of the post-to-pre-collision slip velocity ($\beta = - v_s^{*}/v_s$), ranges between -1 to +1. Generally, $\beta < 0$ in the sliding regime; however, $\beta$ can be positive even if the particles slide during the entire contact. In this appendix, we discuss the condition for $\beta$ to have a positive value in the sliding regime. Also, the DEM simulation of a wall-particle collision shows that despite contact in the gross-sliding regime, $\beta$ can attain a positive value.

For collisions in the sliding regime, $\beta = -1 + (7/2) \mu (1+e_n) (v_{n}/v_{s})$, where $v_n = \vec{v} \cdot \hat{r}$ and $v_s = \vec{v}_s \cdot \hat{k}_s$. For impacts where $(7/2) \mu (1+e_n) (v_n/v_s) > 1$ or $v_{n}/v_{s} > 2/(7\mu(1+e_n))$, $\beta > 0$. For example, if $e_n=0.95$ and $\mu = 0.5$, $\beta>0$ for $(|v_n|/|v_s|)>0.29$. 

\begin{figure}[]
\centering
\includegraphics[scale = 0.25]{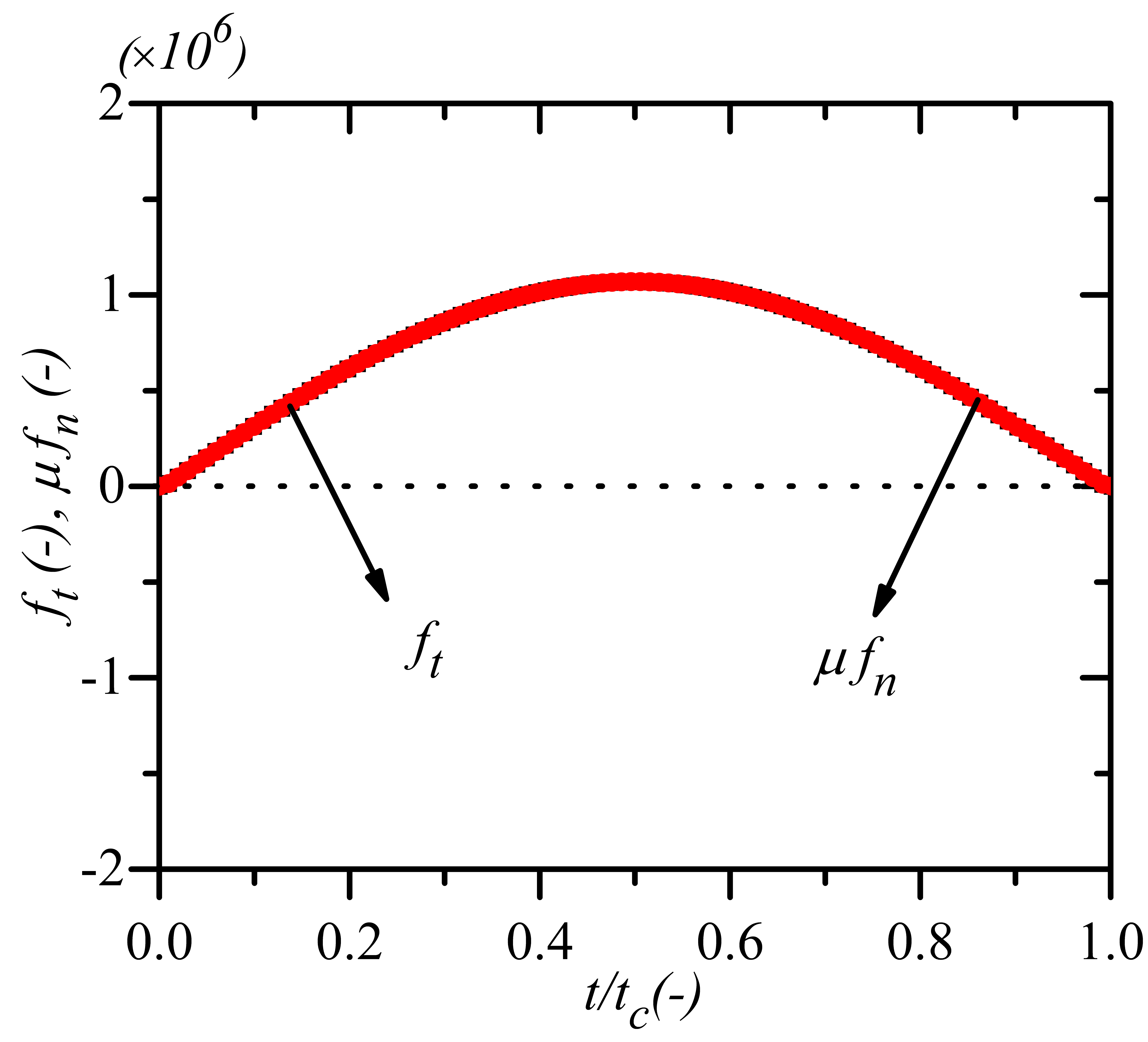}
\caption{ Plot of $f_t$ and $\mu f_n$ obtained from the simulation of a single particle with the wall for contact with $\beta = 0.72$. }\label{fig:ft_mu_fn_single}
\end{figure}

We show an example of a single particle colliding with a wall. Let us consider a particle approaches a wall with linear and angular velocities: $\vec{v} = -100 \hat{i} - 50 \hat{k}$ and $\vec{\omega} = 2.5 \hat{i}$, such that $(|v_n|/|v_s|)=0.5$. 
The velocity and position of the particle are obtained by integrating Newton's equations of motion with an integration time step set at $1/100  t_c$, where $t_c$ is the contact time determined using the linear spring dashpot model. The evolution of tangential and $ \mu$ times the normal force is plotted (Fig. \ref{fig:ft_mu_fn_single}). It is observed that $f_t = \mu f_n$ throughout the entire duration, confirming the gross-sliding.

Pre and post-collision slip velocities are obtained from the simulation to obtain the rotational coefficient of restitution $\beta$. In this case, $\beta = 0.72$, which suggests a slip reversal.

The pre-collision slip velocity of the point of contact of particles, determined by the vector addition of the tangential component of the centre velocity and $\hat{r} \times \vec{\omega}$ ($\vec{v}_s = \vec{v_t} + (d/2)(\hat{r} \times \vec{\omega})$) is $\vec{v}_s = -100 \hat{i} + 6.25 \times 10^{-3} \hat{j} $. In the gross-sliding regime, pre and post-collisional tangential velocity $(\vec{v}_{t} = \vec{v} - \vec{v}_n)$ have the same direction. The direction of $(d/2) (\hat{r} \times \vec{\omega})$ evaluated at pre- and post-collision collision time-steps decides the flipping of the slip of the contact point. The post-collision slip velocity vector $(\vec{v}^{*}_s)$ flips the direction if $\left \| (d/2) \left ( \hat{r} \times \vec{\omega}^{*} \right )\right \| \ge \left \|  \vec{v}_t^{*}\right \|$. In the given example, $\left \|  \vec{v}_t^{*}\right \|$ and $\left \| (d/2)\left ( \hat{r} \times \vec{\omega}^{*} \right )\right \|$ is 50.89 and 122.76, respectively.

\clearpage


\bibliographystyle{apsrev4-1}


\end{document}